\documentclass[12pt]{article}

\usepackage{latexsym}

\usepackage[font=small,labelfont=bf]{caption}

\usepackage{graphics}
\usepackage{graphicx}
\usepackage{epstopdf}
\usepackage{amssymb}
\usepackage{tabularx}
\usepackage{caption}
\usepackage{subcaption}
\usepackage{dcolumn}% Align table columns on decimal point
\usepackage{bm}% bold math
\usepackage{hyperref}
\usepackage{tabularx}
\usepackage{slashbox}
\usepackage{tikz}
\usetikzlibrary{matrix}
\newsavebox{\tempbox}
\usepackage{gensymb}
\newcommand{\be}{\begin{equation}}
\newcommand{\ee}{\end{equation}}
\newcommand{\ba}{\begin{eqnarray}}
\newcommand{\ea}{\end{eqnarray}}
\newcommand{\ban}{\begin{eqnarray*}}
\newcommand{\ean}{\end{eqnarray*}}

\graphicspath{{./Figures/}}
\begin{document}

%\begin{flushright}
%Sofia University\\
%\end{flushright}
%%%%%%%%%%%%%%%%%%%%%%%%%%%%%%%%%%%%%%%%%%%%%%%%%%%%%%%%%%%%%%%%%%%

\title {Image of the Janis-Newman-Winicour naked singularity with a thin accretion disk}

\author{
Galin Gyulchev$^{1}$\footnote{E-mail: \texttt{gyulchev@phys.uni-sofia.bg}}, \, Petya Nedkova$^{1,2}$\footnote{E-mail: \texttt{pnedkova@phys.uni-sofia.bg}}, \, Tsvetan Vetsov$^{1}$\footnote{E-mail: \texttt{vetsov@phys.uni-sofia.bg}},\\ Stoytcho Yazadjiev$^{1,3}$\footnote{E-mail: \texttt{yazad@phys.uni-sofia.bg}}\\ \\
   {\footnotesize${}^{1}$ Faculty of Physics, Sofia University,}\\
  {\footnotesize    5 James Bourchier Boulevard, Sofia~1164, Bulgaria }\\
  {\footnotesize${}^{2}$  Institut f\"{u}r Physik, Universit\"{a}t Oldenburg}\\
  {\footnotesize D-26111 Oldenburg, Germany}\\
  {\footnotesize${}^{3}$ Institute of Mathematics and Informatics,}\\
{\footnotesize Bulgarian Academy of Sciences, Acad. G. Bonchev 8, } \\
  {\footnotesize  Sofia 1113, Bulgaria}}
\date{}
\maketitle

\begin{abstract}
We study the optical appearance and the apparent radiation flux of a thin accretion disk around the static Janis-Newman-Winicour naked singularity. We confine ourselves to the astrophysically most relevant case, when the solution possesses a photon sphere, assuming that the radiation emitted by the disk is described by the Novikov-Thorne model. The observable images resemble closely the visual appearance of the Schwarzschild black hole, as only quantitative differences are present. For the Janis-Newman-Winicour solution the accretion disk appears smaller, and its emission is characterized by a higher peak of the radiation flux. In addition, the most significant part of the radiation is concentrated in a closer neighbourhood of the flux maximum. The results are obtained independently by two alternative methods, consisting of a semi-analytical scheme using the spherical symmetry of the spacetime, and a fully numerical ray-tracing procedure valid for any stationary and axisymmetric spacetime.
\end{abstract}

\section{Introduction}

In the recent years we have witnessed several major experimental  breakthroughs, which have expanded our resources for astrophysical measurements. The gravitational waves were detected, which opened a new channel for obtaining information about the gravitational field. On the other hand, the experiments in the electromagnetic spectrum  achieved a historical success when the Event Horizon collaboration was able to provide the first image of the black hole shadow \cite{EHT01}-\cite{EHT6}. Complemented with further experiments measuring the radiation flux from the accretion disk around the compact objects, and the advances of the astroparticle physics, we gain access to an extremely rich collection of data for examining the properties of the self-gravitating objects in the strong field regime.

General relativity and alternative theories of gravitation predict various compact objects, which arise as particular solutions to the gravitational field equations. However, the possibility of their existence in a real astrophysical environment cannot be determined by purely theoretical arguments. The current experimental resources give the opportunity to test the observable effects, which are characteristic for a certain spacetime, and in this way to differentiate precisely between the possible compact objects. Currently, we possess a solid observational evidence for the existence of black holes in astrophysical scenarios. However, by certain observational phenomena other compact objects can mimic closely the black hole behaviour, and the interpretation of the observational data can remain ambiguous. For example, wormholes and naked singularities, possessing a photon sphere, can give rise to shadow images, which resemble very much that of a black hole \cite{Nedkova:2013}-\cite{Shaikh:2019}.

The current imaging techniques of the compact object environment provide detailed information about the accretion disk surrounding it. Its optical appearance and radiation pattern allow to constrain the physical characteristics of the compact object, such as its mass and angular momentum, and differentiate between the gravitational theories predicting them \cite{Johannsen}-\cite{Takahashi}. Various models of accretion in astrophysically relevant spacetimes are studied by magneto-hydrodynamical simulations in order to provide realistic templates, which can be compared with the experimental results. Yet, in certain cases even a simple physical model can capture the most prominent effects, and provide the basic features, which can observationally distinguish a given spacetime.

In this respect in our work we aim to study a thin accretion disk surrounding the  static Janis-Newman naked singularity \cite{Janis:1968}, and  its optical appearance to a distant observer. If the Einstein field equations are minimally extended by assuming the presence of a massless scalar field, the Janis-Newman solution arises naturally as the unique spherically symmetric solution allowed by the theory. It contains as a certain limit the Schwarzschild black hole, permitting no other spherically symmetric spacetimes with a regular event horizon. Although the existence of naked singularities is prohibited by the cosmic censorship conjecture \cite{Penrose}, there is no conclusive evidence for its validity. Furthermore, naked singularities are formed in a natural way as final stages of the gravitational collapse by assuming suitable initial conditions \cite{Joshi:2011}-\cite{Joshi:2014}, and they are expected to be resolved if quantum gravity effects are taken into account.

Various works investigated the optical properties of the Janis-Newman solution, and the relativistic images, which can arise as a result of the gravitational lensing in this spacetime \cite{Virbhadra:1998}-\cite{Ovgun:2008}. Our aim is to analyze the observable images of the thin accretion disk in its vicinity, and the radiation emitted by it, and to evaluate the possible deviation from the case of the Schwarzschild black hole. In the long-term perspective our results can be applied in the interpretation of the experimental data from the Event Horizon collaboration. Similar studies were performed for the Schwarzschild black hole already in the 70s, when they were considered mainly of academic interest due to the lack of adequate observational resources \cite{Cunningham:1972}-\cite{Fukue:1988}.  With the advance of the visualisation technology much more elaborate images were produced \cite{Muller:2009}-\cite{Muller:2012}. We generalize the semi-analytical procedure developed in \cite{Luminet:1979}, \cite{Muller:2009} for spacetimes of naked singularities, taking advantage of the spherical symmetry of the solution. On the other hand, we implement a fully numerical ray-tracing algorithm, which can be applied for arbitrary stationary and axisymmetric spacetime, and compare the obtained results.

The paper is organized as follows. In section 2 we briefly discuss the Janis-Newman solution focusing on its lensing properties, and the analysis of the stable circular geodesics in the equatorial plane. In section 3 we describe our computational techniques, and present the images, which illustrate the optical appearance of a thin disk, consisting of particles moving on circular equatorial orbits. We interpret the Janis-Newman naked singularity as a possible solution describing the compact object in the center of our galaxy, and scale the observable images accordingly. The images are compared to the Schwarzschild black hole, and some quantitative differences in the visible disk size are evaluated. In section 4 we study the radiation emitted by the disk adopting the Novikov-Thorne thin disk model \cite{Novikov}, \cite{Page:1974}. The apparent radiation flux is illustrated by means of contour plots, and continuous color maps. The characteristic features of the radiation, such as the peak value, and its particular observable distribution are analyzed in comparison to the Schwarzschild solution. In section 5 we summarize our results.

\section{The Janis-Newman-Winicour naked singularity}

The Janis-Newman-Winicour solution is  the most general static and spherically symmetric solution to the Einstein-massless scalar field equations

\begin{eqnarray}
&&{\cal R}_{\mu\nu} = 2\nabla_{\mu}\varphi\nabla_{\nu}\varphi, \\
&&\nabla_{\mu}\nabla^{\mu}\varphi =0. \nonumber
\end{eqnarray}
It is described by the metric \cite{Janis:1968}

\begin{equation}
ds^2 = -\left(1-\frac{b}{r}\right)^{\gamma} dt^2
      +\left(1-\frac{b}{r}\right)^{-\gamma} dr^2
      + \left(1-\frac{b}{r}\right)^{1-\gamma} r^2\left(d\theta^2 + \sin^2\theta d\phi^2\right),
\end{equation}
while the scalar field takes the form
\begin{equation}
\varphi = \frac{q}{b} \ln\left(1-\frac{b}{r}\right).
\end{equation}
The solution is characterized by two real parameters $\gamma$ and $b$, which are related to its $ADM$ mass $M$ and scalar charge $q$ by the expressions
\begin{eqnarray}
\gamma = \frac{2M}{b}, \quad~~~ b =2 \sqrt{M^2+q^2}.
\end{eqnarray}

The parameter $\gamma$ takes the range $0\leq\gamma\leq1$, as in the limit $\gamma=1$ the scalar charge vanishes and the Schwarzschild solution is recovered. For non-trivial scalar fields the solution does not possess an event horizon and describes a naked curvature singularity located at radial distance $r_{cs}=b$.

According to its lensing properties the Janis-Newman solution can be divided into two classes separated by the value of the parameter $\gamma = 0.5$, or equivalently by the scalar charge to mass ratio $q/M = \sqrt{3}$. Weakly naked singularities occurring for $0.5< \gamma< 1$, or $0<q/M < \sqrt{3}$, possess a photon sphere located at $r_{ph} = (2\gamma + 1)b/2$. Consequently,  they are characterized observationally by a shadow, and the structure of the relativistic images typical for them, such as the Einstein rings for example, is similar to that of the Schwarzschild black hole. For $0\leq \gamma< 0.5$ the photon sphere disappears and the light propagation exhibits qualitatively different features. In this regime the solution describes a strongly naked singularity. In the intermediate case $\gamma=0.5$, when the so-called marginally strongly naked singularity is realized there is also no a photon sphere, but the gravitational lensing leads to the appearing of relativistic images. We do not consider specially this situation, because it is qualitatively similar to the weakly naked singularities scenario.

The structure of the timelike circular geodesics in the equatorial plane also depends crucially on the value of the parameter $\gamma$. It is determined by the roots $r_\pm$ of the equation

\begin{equation}\label{ISCO_eq}
r^2\gamma^2 - 2r\gamma(3\gamma + 1) + 2(2\gamma^2 + 3\gamma + 1) =0,
\end{equation}
given explicitly by the expressions

\begin{equation}
r_\pm = \frac{1}{\gamma}\left(3\gamma+1 \pm \sqrt{5\gamma^2 -1}\right).
\end{equation}

Using this equation we can obtain a condition for the stability of the circular orbits. They should possess radial positions belonging to one of the regions  $r>r_{+}$, or $r_{cs}<r<r_{-}$ when $r_{-}$ is larger than the position of the curvature singularity $r_{cs}$, and the second inequality is reasonable. For weakly naked singularities $r_{-} < r_{cs}$ is satisfied and the stable circular orbits occur only for radii larger than the marginally stable orbit $r_{+}$. In this case it represents the innermost stable circular orbit (ISCO), and its position takes values slightly larger than the location of the ISCO for the Schwarzschild solution at $r = 6M$.  For strongly naked singularities two qualitatively different cases are observed. When the solution parameter $\gamma$ takes the range $1/\sqrt{5} < \gamma < 1/2$ we have $r_{-} > r_{cs}$, and the stable circular orbits are represented by two disconnected regions - an annular region for radial distances larger than $r_{+}$, and a circular region surrounding the singularity, which is limited by the marginally stable orbit at $r_{-}$. For $\gamma < 1/\sqrt{5}$ eq. ($\ref{ISCO_eq}$) has no real roots. Therefore, no marginally stable orbits are present, and the stable circular geodesics exist for any radial distance.

\begin{figure*}[t!]
    \centering
    \begin{subfigure}[t]{0.7\textwidth}
        \includegraphics[width=\textwidth]{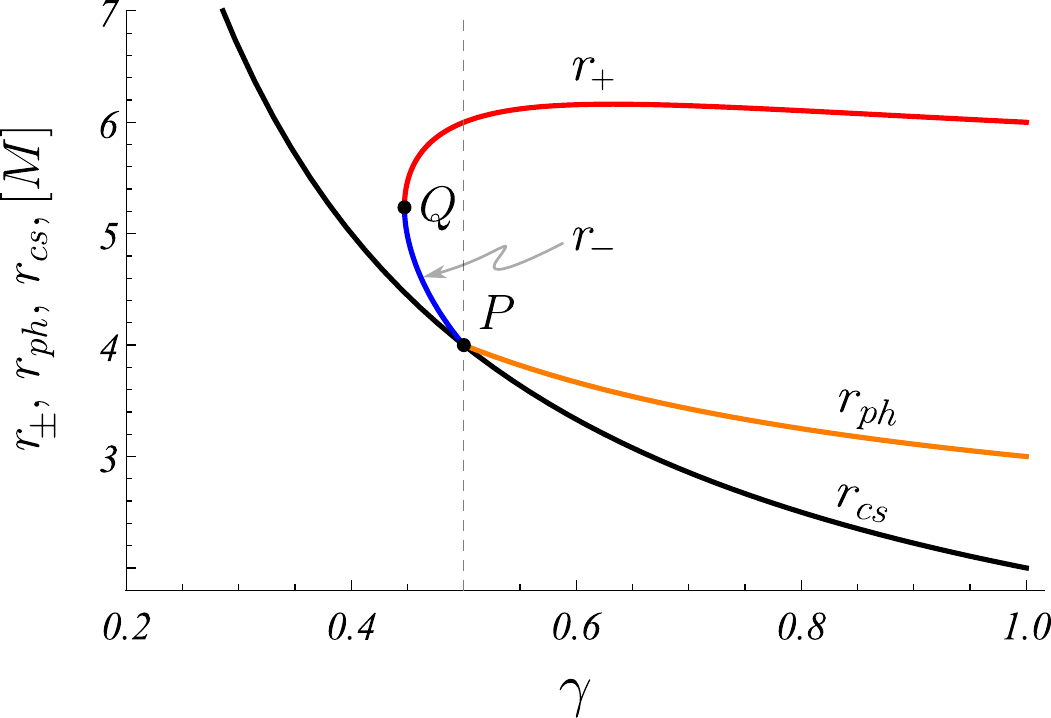}
           \end{subfigure}
           \caption{\label{fig:ISCO}\small Behaviour of the location of the curvature singularity $r_{cs}$ (in black), the photon sphere $r_{ph}$ (in orange), and the outer  and the inner marginally stable orbits $r_{+}$ and $r_{-}$( in red and blue, respectively) when varying the solution parameter $\gamma$. Two critical points are represented, denoted by $P$ and $Q$. At the point $P$ ($\gamma = 1/2$) the position of the photon sphere reaches the curvature singularity and for lower values of $\gamma$  no photon sphere exists. At the same time a second marginally stable orbit appears located at $r= r_{-}$, and the region of stability of the circular geodesics gets disconnected. At the point $Q$ ($\gamma = 1/\sqrt{5}$) the two marginally stable orbits merge, and for lower values of $\gamma$ no marginally stable orbits exist. For $\gamma < 1/\sqrt{5}$  the region of stability of circular geodesics encompasses the whole spacetime.}
\end{figure*}

In fig. $\ref{fig:ISCO}$ we illustrate the behaviour of the location of the curvature singularity  $r_{cs}$, the photon sphere $r_{ph}$, and the marginally stable circular orbits with respect to the value of the parameter $\gamma$. When $\gamma$ decreases, the radial positions of the curvature singularity and the photon sphere increase. However, they grow at a different rate, and for $\gamma = 1/2$ they merge at $r=4M$. At the same time the second marginally stable orbit appears, which delimits the inner circular region of the stable circular geodesics in the vicinity of the singularity. In the range $1/\sqrt{5} < \gamma < 1/2$ the position of the inner marginally stable orbit located at $r_{-}$ increases, while the position of the outer marginally stable orbit at $r_{+}$ decreases, and for $\gamma = 1/\sqrt{5}$ they coincide at $r \approx 5.236$. Then, the two disconnected regions of stability limited by the inner and outer marginally stable orbits merge, and for $\gamma <1/\sqrt{5}$ the stable circular geodesics span the whole spacetime up to the curvature singularity. A detailed analysis of the circular geodesics, however in a different parametrization of the Janis-Newman solution, is performed in \cite{Chowdhury:2012}.

\section{Optical appearance of the Janis-Newman naked singularity with a thin accretion disk}

We consider a simple physical model of a thin accretion disk surrounding the Janis-Newman curvature singularity, which consists of particles moving on stable circular geodesics in the solution spacetime. Each particle is assumed to emit radiation isotropically. The structure of the thin disk depends on the value of the scalar charge to mass ratio, as described in the previous section. In the analysis we restrict ourselves to the case of $q/M < \sqrt{3}$, or $\gamma >1/2$, when a photon sphere is present, and the disk extends from an inner edge determined by the ISCO to infinity. The rest of the parametric space when the disc structure is  more complicated will be investigated in a further work.

The radiation of the particles moving on a circular orbit is deflected by the curved spacetime and as a result the circular orbit will appear distorted to a distant observer. The image of the circular orbits observed at the asymptotic infinity forms the apparent shape of the accretion disk. In practise the disk is represented up to a certain effective radius encompassing the region of strong gravitational field where the most pronounced relativistic effects occur. In our case we investigate the image of the circular geodesics located at radial distances up to the boundary $r = 30M$.

The apparent shape of the accretion disk is calculated independently by two alternative methods comparing the results. Since the spacetime is spherically symmetric, we can develop a semi-analytical procedure taking advantage of the available integrals of motion, which lead to the conservation of the orbital plane, and the photon's energy $E$ and angular momentum $L$ on the geodesic trajectory. We consider the constraint equation for the null geodesics $g_{\mu\nu}{\dot x}^{\mu}{\dot x}^{\nu} =0$, where the dot denotes differentiation with respect to an affine parameter. For the Janis-Newman naked singularity it takes the explicit form

\begin{equation}
{\dot r}^2 + \frac{L^2}{r^2}\left(1-\frac{b}{r}\right)^{2\gamma -1}=E^2.
\end{equation}

Using the geodesic equation for the azimuthal angle

\begin{equation}
{\dot \phi}= \frac{L}{r^2}\left(1-\frac{b}{r}\right)^{\gamma -1},
\end{equation}
we can  obtain its variation along the photon trajectory as a function of the radial distance

\begin{equation}\label{delta_phi_int}
\Delta\phi = \int{\frac{dr}{r^2\left(1-\frac{b}{r}\right)^{1-\gamma}\sqrt{\frac{1}{D^2} -\frac{1}{r^2}\left(1-\frac{b}{r}\right)^{2\gamma -1}}}}.
\end{equation}

The azimuthal angle depends on a single impact parameter $D=L/E$ determined by the ratio of the angular momentum and the energy on the geodesic. In our case we are interested in the photon trajectories, which are emitted by a particle orbiting at a certain radial distance $r_0$ in the disk, and after scattering by the gravitational field of the Janis-Newman solution, reach a distant observer. Therefore, the integration is performed numerically from the orbit's location to infinity. In the computations we assume that the asymptotic infinity corresponds effectively to a radial distance of the observer $r_{obs}=5000M$.  The photon trajectory is projected on the observer's sky giving rise to an image determined by two celestial coordinates $\xi$ and $\eta$. If we consider the general form of a static spherically symmetric metric in spherical coordinates
\begin{equation}\label{metric_st}
ds^2=g_{tt}\,dt^2+g_{\phi\phi}\,d\phi^2+g_{rr}\,dr^2+g_{\theta\theta}\,d\theta^2\,,
\end{equation}
we can define a local reference frame at the observer's location with basis vectors

\begin{eqnarray}
\hat{e}_{(\theta)}&=&\frac{1}{\sqrt{g_{\theta\theta}}}\partial_\theta,\quad~~~ \hat{e}_{(r)}=\frac{1}{\sqrt{g_{rr}}}\partial_r,\\
\hat{e}_{(\phi)}&=&\frac{1}{\sqrt{g_{\phi\phi}}}\partial_\phi, \qquad~~~\hat{e}_{(t)}=\frac{1}{\sqrt{-g_{tt}}}\,\partial_t, \nonumber
\end{eqnarray}
expanded in the coordinate basis $\{\partial_t,\partial_r,\partial_\theta,\partial_\phi\}$. The radial vector $\hat{e}_{(r)}$ is assumed to point into the direction of the compact object. In this local tetrad the celestial coordinate $\eta$ is defined as the angle between the null geodesic plane and the  basis vector $\hat{e}_{(\phi)}$, while $\xi$ is the angle between the trajectory and the compact object direction $\hat{e}_{(r)}$. Due to the spherical symmetry we can express the variation of the azimuthal angle along the geodesic by means of the celestial angle $\eta$, and the inclination angle $i$ of the observer with respect to the orbital plane of the emitting particle \cite{Muller:2009}

\begin{eqnarray}\label{delta_phi}
\cos\phi = - \frac{\sin\eta\tan i}{\sqrt{\sin^2\eta\tan^2 i + 1}}, \quad~~~ \sin\phi = \frac{1}{\sqrt{\sin^2\eta\tan^2 i + 1}}.
\end{eqnarray}

On the other hand, the celestial angle $\xi$ is connected to the impact parameter $D$ on the geodesic as

\begin{equation}\label{xi}
\xi =\arcsin{\frac{D}{r_{obs}}\left(1 - \frac{b}{r_{obs}}\right)^{\gamma - \frac{1}{2}}}.
\end{equation}

For a given radial position $r_s$ of the orbit of the emitting particle and a given inclination angle $i$ of the observer  eqs. ($\ref{delta_phi_int}$) and ($\ref{delta_phi}$) define a relation between the impact parameter of the photon trajectory and the corresponding observational angle $\eta$.

\begin{equation}\label{eta_b}
\int^{r_{obs}}_{r_{s}}{\frac{dr}{r^2\left(1-\frac{b}{r}\right)^{1-\gamma}\sqrt{\frac{1}{D^2} -\frac{1}{r^2}\left(1-\frac{b}{r}\right)^{2\gamma -1}}}} = -\arccos{ \frac{\sin\eta\tan i}{\sqrt{\sin^2\eta\tan^2 i + 1}}}
\end{equation}

Using eq. ($\ref{xi}$) it is transformed into a relation between the celestial coordinates $\xi= \xi(\eta)$. In practise, by scanning all the observational angles $\eta \in [0,2\pi]$ we can obtain  the possible impact parameters $D$, which arise as solutions of ($\ref{eta_b}$) for a particular inclinational angle $i$ and initial position $r_s$. Then, the curve $\xi= \xi(\eta)$ determined by ($\ref{xi}$) defines the observable image of the circular orbit located at radial distance $r_s$.

The variation of the azimuthal angle ($\ref{delta_phi}$) is supposed to belong to the range $\Delta\phi \in [0, \pi)$. However, certain trajectories are deflected at larger angles, or perform several circles around the compact object before reaching the observer. In order to take these cases into account eq. $\ref{delta_phi_int}$ should be generalized to

\begin{equation}\label{eta_b}
\int^{r_{obs}}_{r_{s}}{\frac{dr}{r^2\left(1-\frac{b}{r}\right)^{1-\gamma}\sqrt{\frac{1}{D^2} -\frac{1}{r^2}\left(1-\frac{b}{r}\right)^{2\gamma -1}}}} = k\pi -\arccos{ \frac{\sin\eta\tan i}{\sqrt{\sin^2\eta\tan^2 i + 1}}},
\end{equation}

\noindent
where $k$ is a non-negative integer. Trajectories with $k=0$ give rise to direct images, while those with $k=1$ form  secondary images. Null geodesics with $k\geq2$ result in images very close to the image of the photon sphere corresponding to the rim of the shadow.

We calculate the image of the circular orbits independently in a second way by using a fully numerical ray-tracing procedure. The computation scheme does not take advantage of the spherical symmetry of the solution, and it can be applied for any stationary and axisymmetric spacetime. The null geodesic equations are represented in Hamiltonian form and integrated numerically. A photon trajectory reaching a distant observer is projected on their observational plane by means of two angles $\alpha$ and $\beta$ related to the photon's 4-momentum as \cite{Cunha:2016a}

\begin{eqnarray}\label{p_alpha}
&&p_\theta=\sqrt{g_{\theta\theta}}\sin\alpha,\qquad\qquad\,\,\, p_\phi =L =\sqrt{g_{\phi\phi}}\sin\beta\,\cos\alpha, \nonumber \\[2mm]
&&p_r=\sqrt{g_{rr}}\cos\beta\,\cos\alpha,\qquad p_t = -E.
\end{eqnarray}

Given the radial position and the inclination angle of the observer we can use the angles $\alpha$ and $\beta$ as initial data on the geodesic and integrate the trajectory backwards to the position of the emitting particle, which moves on a circular orbit in the equatorial plane. Scanning all the observational angles in the range $\alpha \in [0,\pi]$ and $\beta \in [-\pi/2,\pi/2]$ we can select the appropriate ones which correspond to trajectories emitted by the accretion disk. In practise, they should represent solutions to the geodesic equations passing through a point with coordinates $\theta = \pi/2$ and $r \in [r_{ISCO}, r_{out}]$, where $r_{ISCO}$ is the position of the innermost stable circular orbit and $r_{out}$ is some outer boundary, which we select by physical reasons. This set of observational angles defines the  image of the accretion disk on the observer's sky \footnote{For an asymptotic observer the celestial coordinates $\alpha$ and $\beta$ are related to the observational angles $\eta$ and $\xi$ introduced in the semi-analytical computation procedure as $\alpha = \xi\cos\eta$, $\beta = \xi\sin\eta$.}.

Using the described procedures we compute the observable image of a thin accretion disk around Janis-Newman naked singularity in the region of the parametric space $\gamma \in (0.5, 1)$ when a photon sphere is present. We compare the optical appearance of the accretion disk around the naked singularity with the case of the Schwarzschild black hole, which corresponds  to the limit $\gamma= 1$. In order to illustrate the effects of the absence of an event horizon most clearly we present the image of the Janis-Nemwan solution for a value of the parameter $\gamma$ close to the lower limit $\gamma = 0.5$, when the deviation from the Schwarzschild black hole is most pronounced. The optical appearance of the accretion disk depends substantially on the  inclination angle of the observer, therefore we provide two sets of images for an observer with inclination angle  $i= 60^\circ$, and a nearly equatorial observer with $i= 80^\circ$. The observer is located at the radial distance $r_{obs}=5000M$, which corresponds effectively to the asymptotic infinity. Modeling the massive dark object Sgr A* in the center of our Galaxy as a gravitational lens we present the celestial coordinates of the images ($\alpha,\beta$) in terms of $\mu$as. We take into account that the observer is positioned at distance $D_{OL}=8.33$ kpc from the lens. In this circumstances the scaling factor between the radial and physical distances is $r_{obs}/D_{OL}\approx1.24\times10^{-7}$. According to \cite{Gillessen:2009} the lens mass is $M=4.31\times10^{6}M_{\odot}$, so $M/D_{OL}\approx2.47\times10^{-11}$.

\begin{figure}[h!]
    		\setlength{\tabcolsep}{ 0 pt }{\footnotesize\tt
		\begin{tabular}{ cc }
           \includegraphics[width=0.5\textwidth]{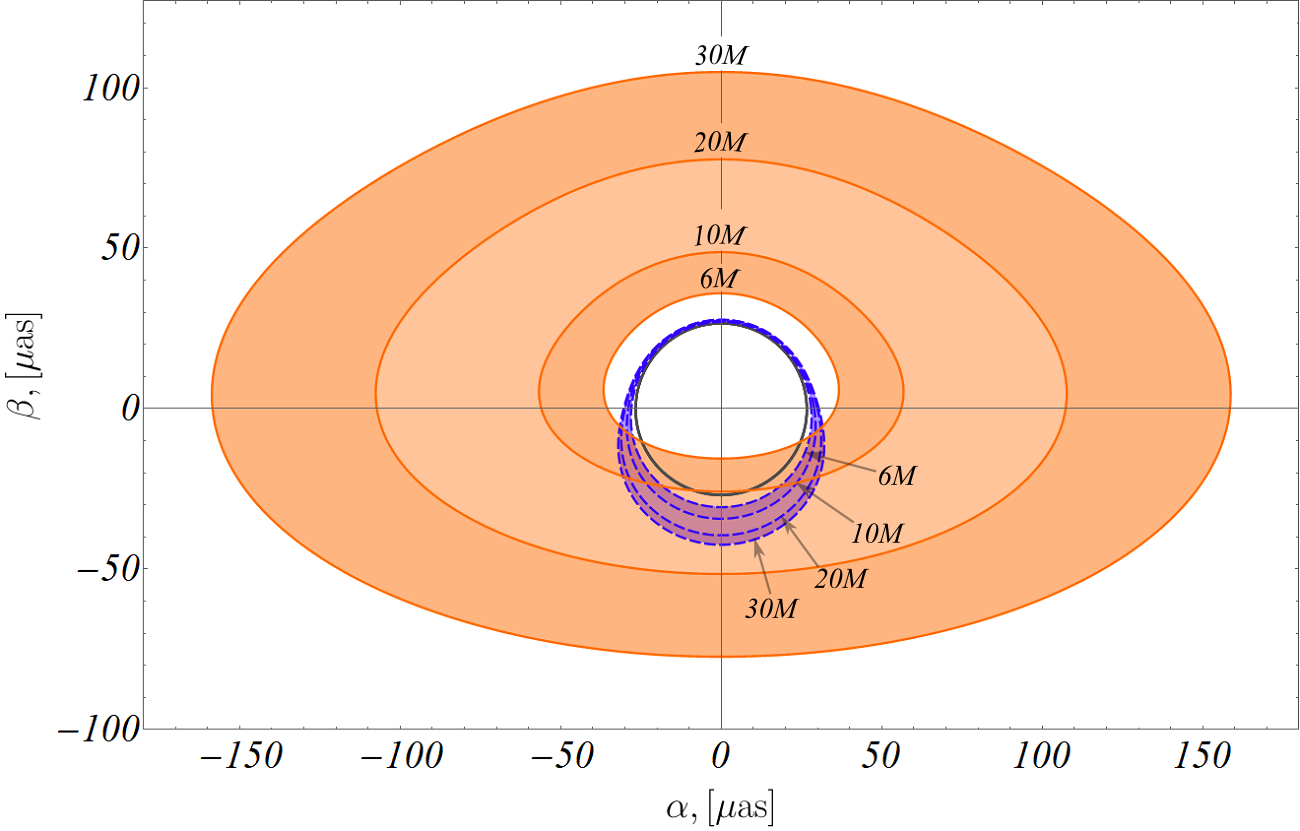}
             \includegraphics[width=0.5\textwidth]{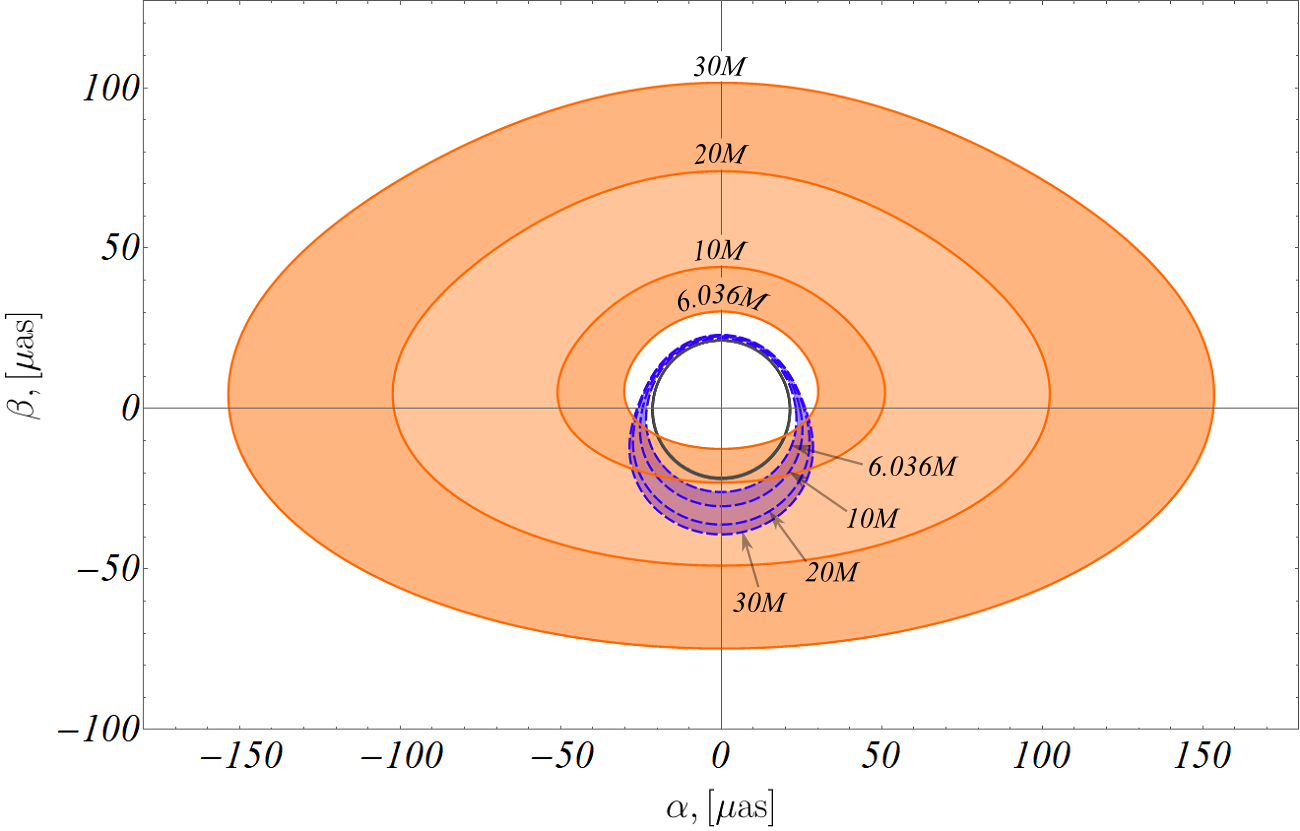} \\[1mm]
           		\end{tabular}}
 \caption{\label{fig:IsoR1}\small Optical appearance of the thin accretion disk around Schwazrschild black hole (left) and Janis-Newman naked singularity (right) with solution parameter $\gamma = 0.51$. The inclination angle of the observer is $i=60^\circ$, and their radial position is $r=5000M$. The direct image of the accretion disk is depicted in orange, while the secondary image is in blue. The black circle represents the optical appearance of the photon sphere, i.e. the boundary of the shadow of the compact object. We present the image of the accretion disk in the region between the ISCO and an outer circular orbit located at $r=30M$, which illustrates the regime of strong gravitational field. The innermost stable circular orbit is located at $r=6M$ for the Schwarzschild solution, and at $r=6.03$ for the Janis-Newman naked singularity with $\gamma= 0.51$. }
\end{figure}

\begin{figure}[h!]
    		\setlength{\tabcolsep}{ 0 pt }{\footnotesize\tt
		\begin{tabular}{ cc }
           \includegraphics[width=0.5\textwidth]{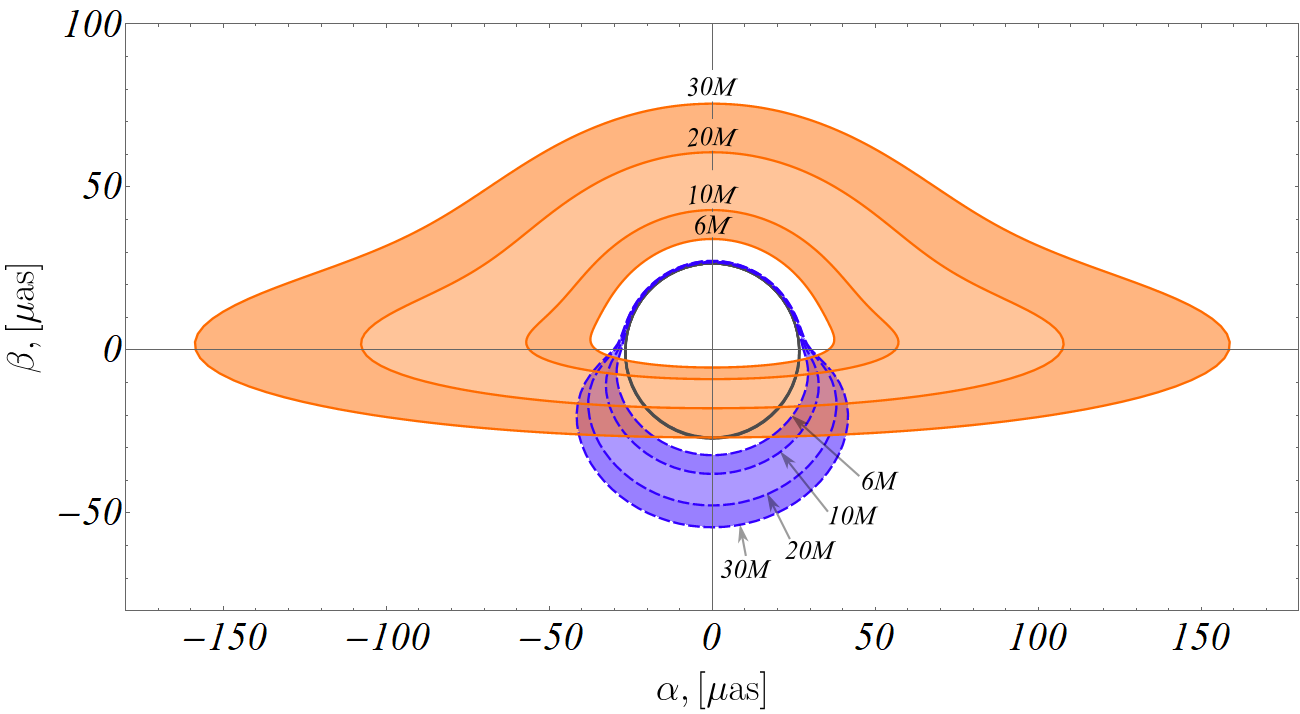}
             \includegraphics[width=0.5\textwidth]{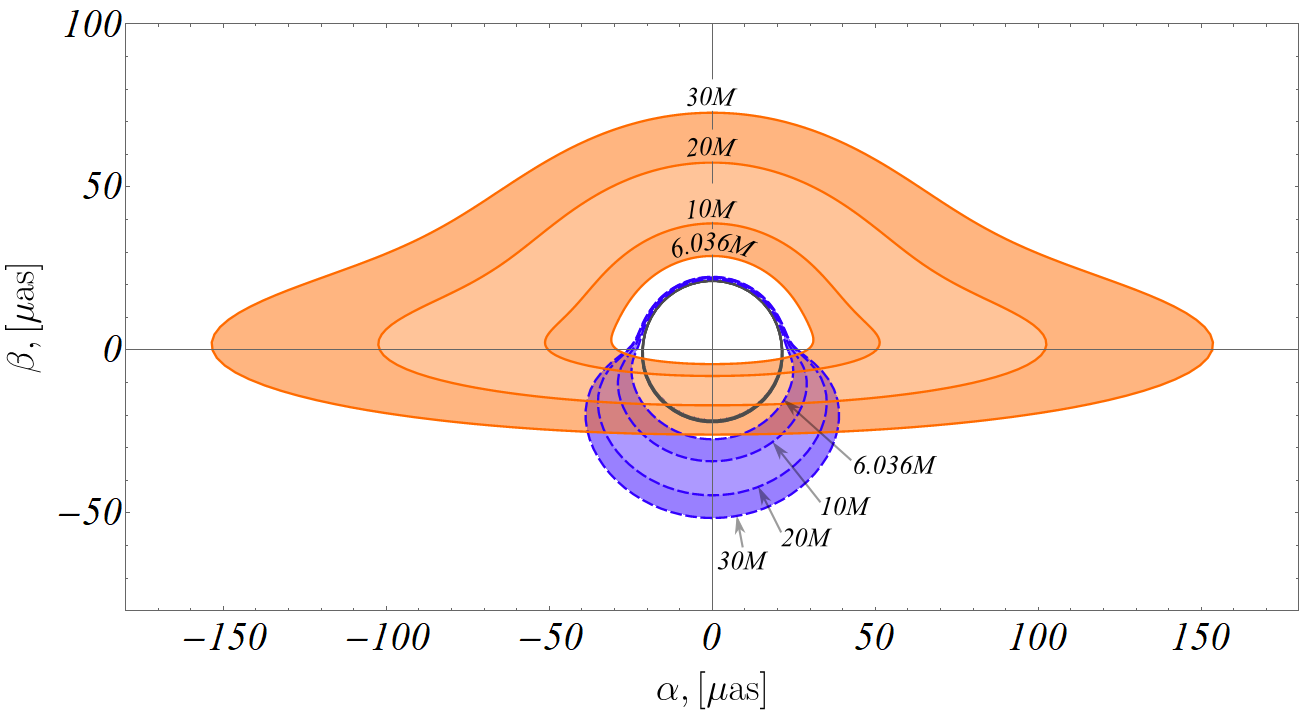} \\[1mm]
           		\end{tabular}}
 \caption{\label{fig:IsoR2}\small Optical appearance of the thin accretion disk around Schwazrschild black hole (left) and Janis-Newman naked singularity (right) with solution parameter $\gamma = 0.51$, and an inclination angle of the observer $i=80^\circ$. We use the same conventions as in fig. $\ref{fig:IsoR1}$. }
\end{figure}

The optical appearance of the accretion disk is presented in figs. $\ref{fig:IsoR1}$-$\ref{fig:IsoR2}$, illustrating the image of the circular orbits in the equatorial plane located between the ISCO and an outer orbit with radius $r=30M$. In each figure two images are superposed. The first one is generated by the numerical procedure and represents the visual appearance of a continuous distribution of circular orbits with radii belonging to the range $r_{ISCO}\leq r\leq 30M$. The second image is computed by the semi-analytical procedure and illustrates the optical appearance of a discrete set of circular orbits corresponding to particular radial distances. The visualization of these orbits is represented by solid lines in the figures, where we also specify the radial position of each orbit. We can further differentiate between direct images, characterized by variation of the azimuthal angle along the photon trajectory in the range $\Delta\phi \in [0, \pi)$, and secondary images for which the null geodesic is deflected at a larger angle or circles several time around the compact object before reaching the observer, i.e.  $\Delta\phi \in [k\pi, (k+1)\pi], k\in {\cal N}$. The direct images are generated by photons emitted in direction above the equatorial plane, while secondary images with $k=1$ correspond to photons heading in direction below the equatorial plane. These two cases are depicted in orange and blue, respectively. We further represent in black the apparent shape of the photon sphere, which outlines the boundary of the shadow of the compact object. The photons which perform several circles around the compact object before reaching the observer ($k\geq2$) possess impact parameters very close to that of the photon sphere. Therefore, they give rise to images, which approach the shadow rim, and practically cannot be distinguished in the figures.

The thin accretion disks around the Schwarzschild black hole and a weakly naked Janis-Newman singularity appear in a very similar way to a distant observer. The two images share the same qualitative features and can be hardly distinguished by observations. Quantitatively they can be differentiated by evaluating the size of the apparent shape of the disk. The Janis-Newman singularity possesses a stronger focusing effect on the photon trajectories, since the circular orbits result in a smaller image than the corresponding ones for the Schwarzschild solution. We investigate the shrinking of the visual size by comparing the ranges of the celestial coordinates in the horizontal and vertical directions appearing in the images of circular orbits with certain radii in both spacetimes. As a quantitative measure for the horizontal shrinking we take the difference $\Delta\alpha$ between the maximal values of the $\alpha$-coordinate in the image of a particular orbit for the Schwarzschild and the Janis-Newman spacetimes, $\alpha_{Schw}$ and $\alpha_{JN}$ respectively, normalized by its maximal value for the Schwarzschild solution

\begin{equation}
\frac{\Delta\alpha}{\alpha} = \frac{\alpha_{Schw} - \alpha_{JN}}{\alpha_{Schw}}.
\end{equation}

In a similar way we can estimate the vertical shrinking by the normalized difference of the $\beta$-coordinate for the two solutions. We take the difference between the maximum value $\beta^{+}$ and the minimum value $\beta^{-}$  of the $\beta$-coordinate in each image for the Schwarzschild and Janis-Newman solutions, i.e. $\beta_{Schw} = \beta^{+}_{Schw} - \beta^{-}_{Schw}$ and $\beta_{JN} = \beta^{+}_{JN} - \beta^{-}_{JN}$. Then, we define a measure of the variation in the $\beta$-direction as

\begin{equation}
\frac{\Delta\beta}{\beta} = \frac{\beta_{Schw} - \beta_{JN}}{\beta_{Schw}}.
\end{equation}

The results are presented in tables  $\ref{alpha}$-$\ref{beta}$ for the inclination angles $i = 60^\circ$ and $i = 80^\circ$, examining several circular orbits. The reduction of the visual size of the orbit depends on its radius and on the order of the image $k$. For a given image order the innermost stable circular orbit exhibits the largest relative shrinking, and the effect decreases when the radius of the orbit grows. For most of the orbits the reduction of their visual size increases with the increase of the image order. This effect is consistent with the fact that for images of higher order we observe a smaller relative distance between the visual positions of the different orbits, so that all of them appear closer to the ISCO. However, the effect is violated for orbits in the vicinity of the ISCO, for which the reduction of the visual size for the secondary image with order $k=1$ is less than the reduction for the direct image with $k=0$ (see table $\ref{alpha}$-$\ref{beta}$).

The shrinking of the visual size was observed previously for the image of the photon sphere of the Janis-Newman solution \cite{Virbhadra:2002}. Decreasing the solution parameter $\gamma$ in the range $\gamma\in(0.5, 1]$, or equivalently increasing the scalar charge to mass ratio, the radial position of the photon sphere increases. At the same time we observe reduction of its apparent size on the observer's sky. In this case the phenomenon can be explained analytically by calculating the impact parameter on the photon sphere, which determines the size of its observable image, and showing that it is a decreasing function of the parameter $\gamma$.

%\begin{table}[h!]
%\caption{\label{alpha}The relative visual size shrinking in the horizontal direction $\frac{\Delta \alpha}{\alpha}[\%]$ of the circular equatorial orbits for $\gamma=0.51$ with %respect to the corresponding orbits for the Schwarzschild black hole for different image order $k$.}
%\centering
%\begin{tabular}{|l|*{4}{c|}}\hline
%\backslashbox{$k$}{$r/M$}
%&\makebox[3em]{6.03}&\makebox[3em]{10}&\makebox[3em]{20}
%&\makebox[3em]{30}\\\hline\hline
%0 &17.5\% &10.1\%&4.9\%&3.2\%\\\hline
%1 &15.6\%&11.3\%&7.7\%&6.7\%\\\hline
%2 &19.9\%&19.8\%&19.6\%&19.5\%\\\hline
%\end{tabular}
%\end{table}
%
%\begin{table} [h!]
%\caption{\label{beta}The relative visual size shrinking in the vertical direction $\frac{\Delta \beta}{\beta}[\%]$ of the circular equatorial orbits for $\gamma=0.51$ with %respect to the corresponding orbits for the Schwarzschild black hole for different image order $k$.}
%\centering
%\begin{tabular}{|l|*{4}{c|}}\hline
%\backslashbox{$k$}{$r/M$}
%&\makebox[3em]{6.03}&\makebox[3em]{10}&\makebox[3em]{20}&\makebox[3em]{30}\\\hline\hline
%0 &16.1\% &9.7\%&5.2\%&3.6\%\\\hline
%1 &17.0\%&13.6\%&10.7\%&9.5\%\\\hline
%2 &19.9\%&19.6\%&19.4\%&19.4\%\\\hline
%\end{tabular}
%\end{table}

\begin{table}
\caption{\label{alpha} \small The relative visual size shrinking in the horizontal direction $\frac{\Delta \alpha}{\alpha}[\%]$ of the circular
equatorial orbits for $\gamma=0.51$ with respect to the corresponding orbits for the Schwarzschild
black hole for different image order $k$. We study two inclination angles $i=60^{\circ}$ (a), and $i=80^{\circ}$ (b).}
\centering
\small
\subcaption{\small $i=60^{\circ}$}
\begin{tabular}{|l|*{7}{c|}}\hline
\backslashbox{$k$}{$r/M$}
&\makebox[3em]{6.03}&\makebox[3em]{7}&\makebox[3em]{8}&\makebox[3em]{9}&\makebox[3em]{10}&\makebox[3em]{20}&\makebox[3em]{30}\\\hline\hline
0 &17.99\% &15.15\% &12.98\% &11.43\% &10.18\% &4.91\%  &3.24\%\\\hline
1 &16.84\% &15.86\% &15.11\% &14.51\% &14.02\% &11.70\% &10.85\%\\\hline
2 &19.91\% &19.86\% &19.82\% &19.78\% &19.76\% &19.66\% &19.63\%\\\hline
\end{tabular}

\vspace{4mm}

\subcaption{\small $i=80^{\circ}$}
\begin{tabular}{|l|*{7}{c|}}\hline
\backslashbox{$k$}{$r/M$}
&\makebox[3em]{6.03}&\makebox[3em]{7}&\makebox[3em]{8}&\makebox[3em]{9}&\makebox[3em]{10}&\makebox[3em]{20}&\makebox[3em]{30}\\\hline\hline
0 &17.51\% &14.93\% &12.80\% &11.35\% &10.09\%&4.91\%&3.23\%\\\hline
1 &15.61\% &14.16\% &12.91\% &11.99\% &11.33\% &7.72\% &6.66\%\\\hline
2 &19.91\% &19.82\% &19.75\% &19.70\% &19.76\% &19.62\% &19.53\%\\\hline
\end{tabular}
\end{table}

\begin{table}
\caption{\label{beta} \small The relative visual size shrinking in the vertical direction $\frac{\Delta \beta}{\beta}[\%]$ of the circular
equatorial orbits for $\gamma=0.51$ with respect to the corresponding orbits for the Schwarzschild
black hole for different image order $k$. We study two inclination angles $i=60^{\circ}$ (a) and $i=80^{\circ}$ (b).}
\centering
\small
\subcaption{\small $i=60^{\circ}$}
\begin{tabular}{|l|*{7}{c|}}\hline
\backslashbox{$k$}{$r/M$}
&\makebox[3em]{6.03}&\makebox[3em]{7}&\makebox[3em]{8}&\makebox[3em]{9}&\makebox[3em]{10}&\makebox[3em]{20}&\makebox[3em]{30}\\\hline\hline
0 &17.25\% &14.52\% &12.53\% &11.04\% &9.89\%&4.88\%&3.24\%\\\hline
1 &17.00\% &16.00\% &15.26\% &14.70\% &14.24\%&12.15\%&11.40\%\\\hline
2 &19.80\% &19.72\% &19.66\% &19.62\% &19.58\%&19.45\%&19.41\%\\\hline
\end{tabular}

\vspace{4mm}

\subcaption{\small $i=80^{\circ}$}
\begin{tabular}{|l|*{7}{c|}}\hline
\backslashbox{$k$}{$r/M$}
&\makebox[3em]{6.03}&\makebox[3em]{7}&\makebox[3em]{8}&\makebox[3em]{9}&\makebox[3em]{10}&\makebox[3em]{20}&\makebox[3em]{30}\\\hline\hline
0 &16.11\% &13.78\% &12.06\% &10.77\% &9.75\%&5.18\%&3.58\%\\\hline
1 &17.00\% &15.81\% &14.91\% &14.20\% &13.62\% &10.70\% &9.47\%\\\hline
2 &19.86\% &19.60\% &19.53\% &19.48\% &19.60\% &19.43\% &19.38\%\\\hline
\end{tabular}
\end{table}

\begin{figure}[t!]
    		\setlength{\tabcolsep}{ 0 pt }{\small\tt
		\begin{tabular}{ cc}
           \includegraphics[width=0.9\textwidth]{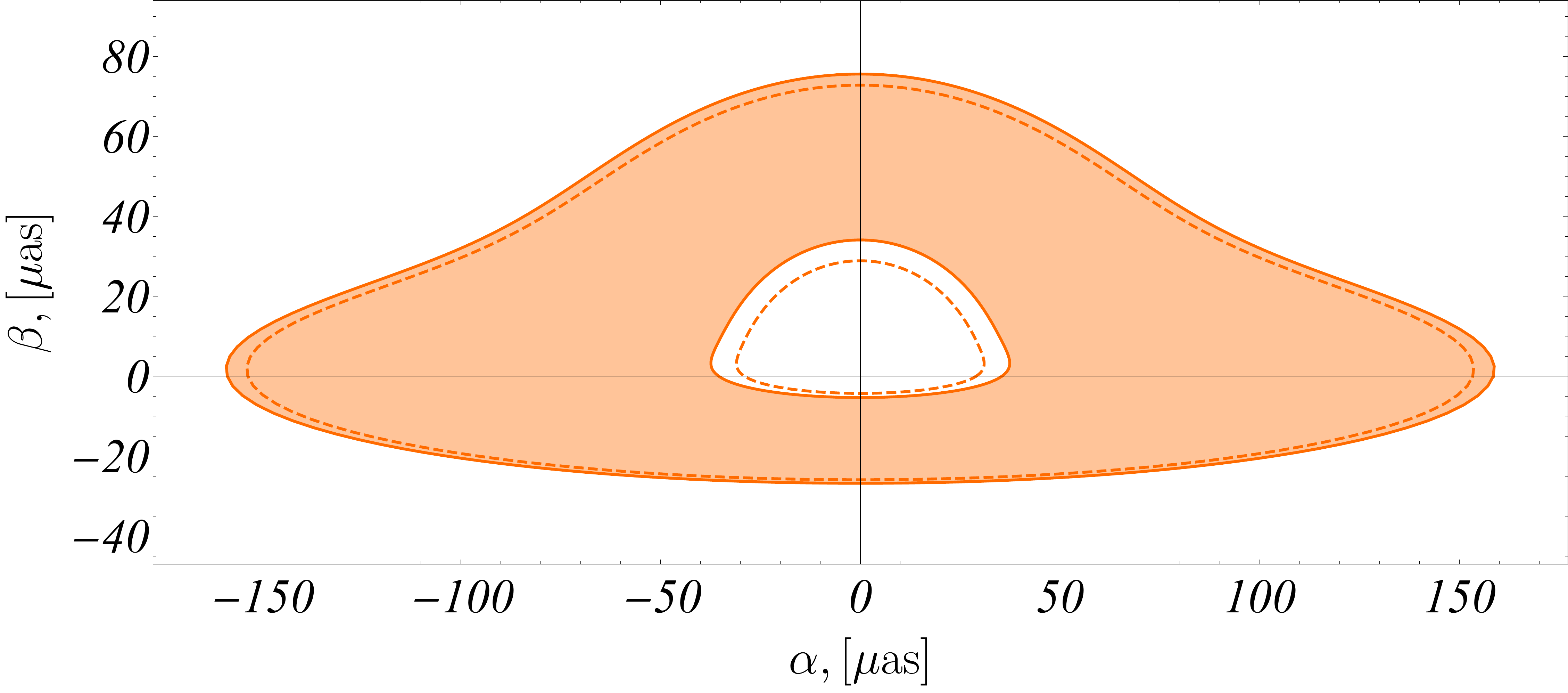} \\[1mm]
           \hspace{1cm} $k=0,\,\,\,\delta A=4.75\%$ \\[5mm]
           \includegraphics[width=0.39\textwidth]{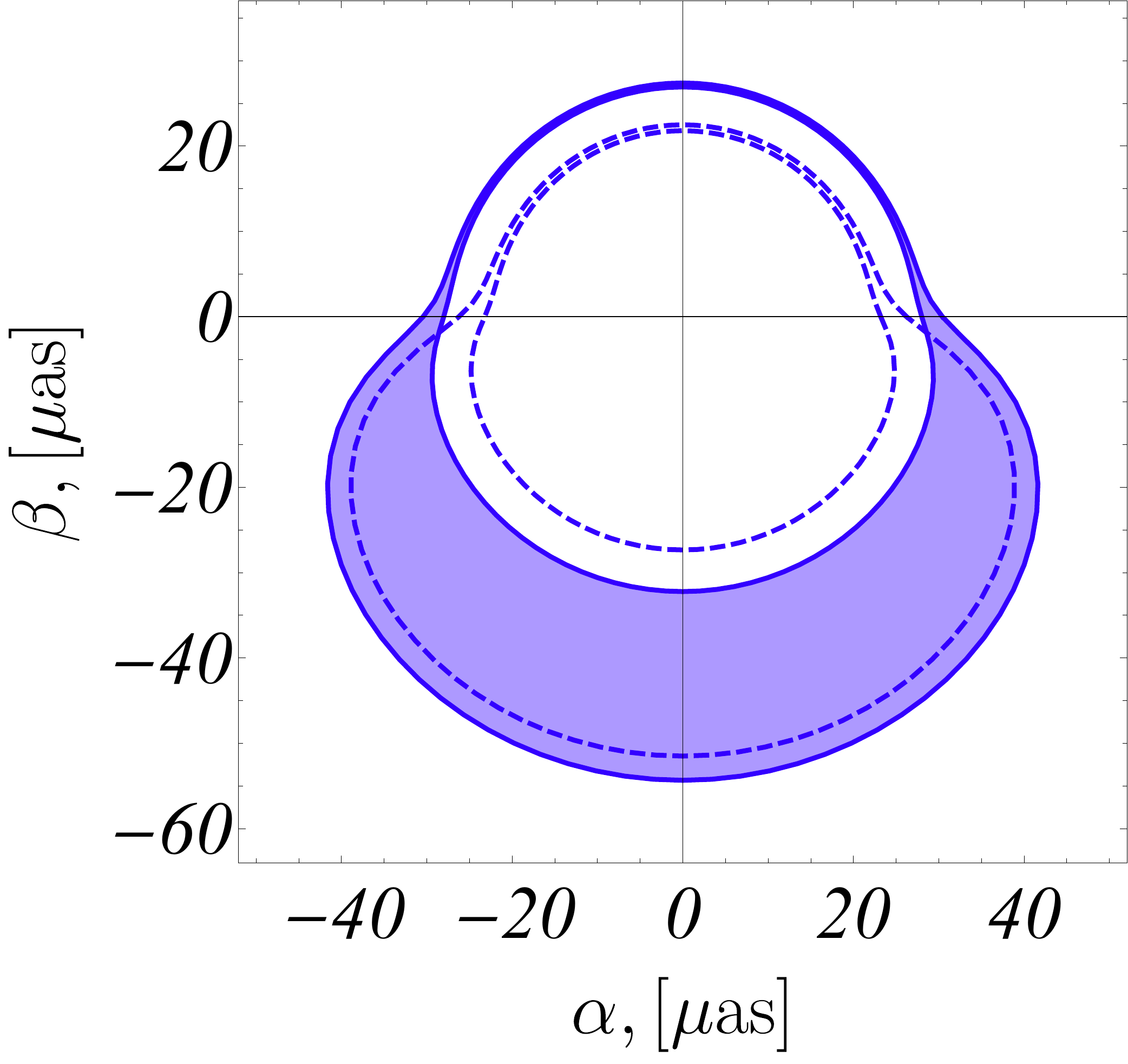}\hspace{1.5cm}
		   \includegraphics[width=0.39\textwidth]{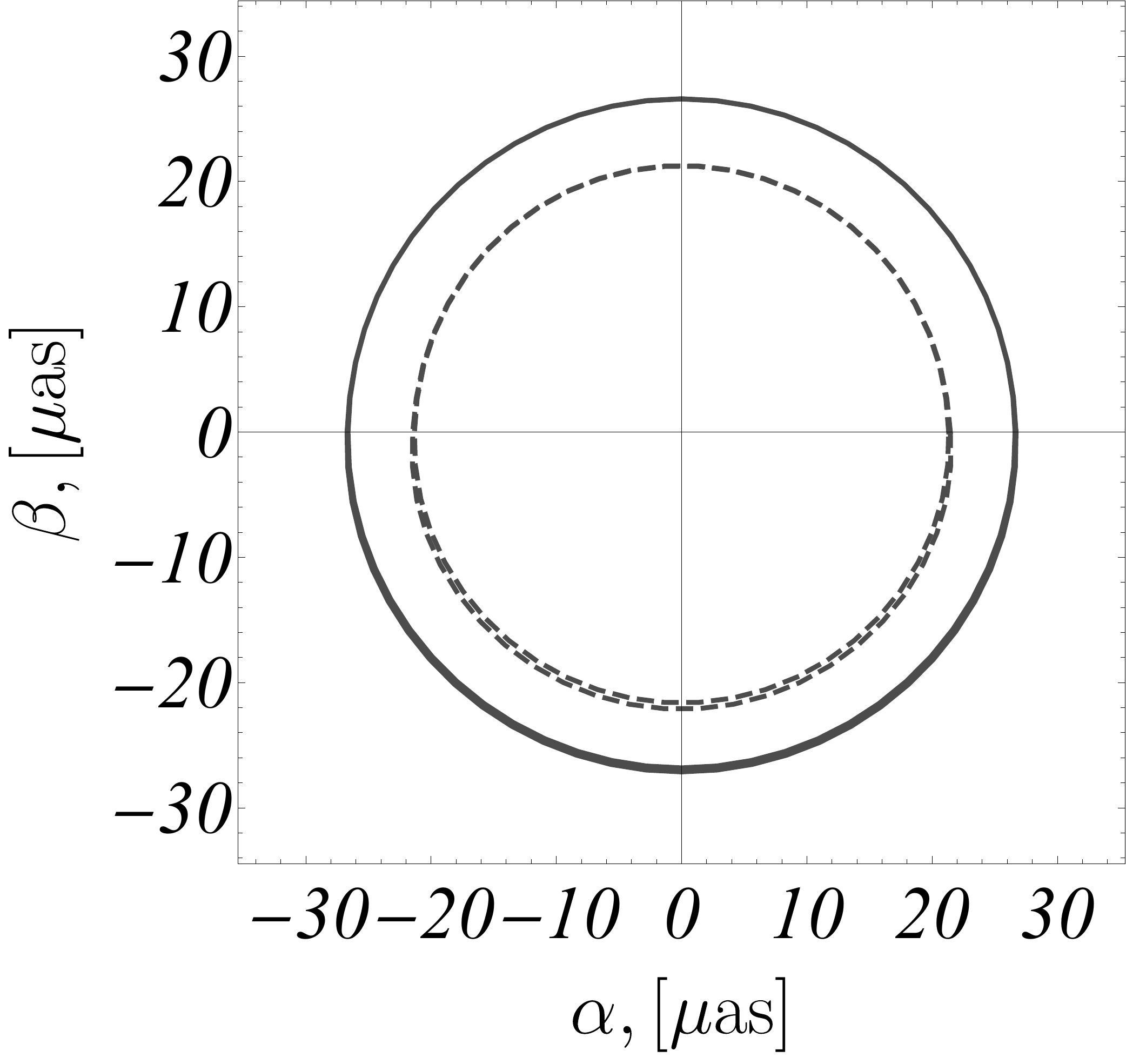} \\[1mm]
           \hspace{1cm}  $k=1,\,\,\,\delta A=-2.17\%$ \hspace{3.8cm}  $k=2,\,\,\,\delta A=-36.24\%$
        \end{tabular}}
 \caption{\label{fig:Area80deg}\small The relative visual area size shrinking $\delta A$ for the Janis-Newman naked singularity with solution parameter $\gamma = 0.51$ (dashed line) with respect to the corresponding disk region for the Schwarzschild black hole (solid line) for different image order $k=0,1,2$. The inclination angle is $i=80^{\circ}$. For secondary images with $k=1,2$ the area bounded between the image of the ISCO and the outer boundary of the disk at $r=30M$ is greater for the Janis-Newman solution than the corresponding area for the Schwarzschild black hole.}
\end{figure}

We can further study the influence of the reduction of the visual orbital size on the apparent area of the accretion disk. We calculate the area of the disk image bounded between the image of the ISCO and the outer circular orbit at $r=30M$ for the Janis-Newman solution with $\gamma=0.51$ and for the Schwarzschild black hole. Denoting these quantities by  $A_{JN}$ and $A_{Schw}$ respectively, we define a measure of the relative visual area deviation between the two solutions

\begin{equation}
\delta A=\frac{A_{Schw}-A_{JN}}{A_{Schw}}.
\end{equation}
In table $\ref{area}$ we present the values of $\delta A$ for the direct image ($k=0$) and the secondary images ($k=1,2$) for two inclination angles $i=60^{\circ}$ and $i=80^{\circ}$, while the corresponding areas are visualized in fig. \ref{fig:Area80deg}. The Schwarzschild black hole leads to larger direct images of the accretion disk than the Janis-Newman naked singularity, while for secondary images we observe the opposite effect. This scenario is represented by negative values of the quantity $\delta A$ in table $\ref{area}$.

%\savebox{\tempbox}{\begin{tabular}{@{}r@{}l@{\space}}
			%&k\\$i$
		%\end{tabular}}

%		\begin{table}
%         \centering
			%\caption{\label{area}The relative visual area size shrinking $\delta A$ for the Janis-Newman naked singularity with solution parameter $\gamma = 0.51$ with respect %to the corresponding disk region for the Schwarzschild black hole for different image order $k=0,1,2$. We study two inclination angles  $i=60^{\circ}$ and $ i=80^{\circ}$.}
			%\[\begin{array}{|l|*{3}{c|}}\hline
			%\tikz[overlay]{\draw (0pt,\ht\tempbox) -- (\wd\tempbox,-\dp\tempbox);}%
			%\usebox{\tempbox}\hspace{\dimexpr 1pt-\tabcolsep}
			%& 0 & 1 & 2 \\
			%\hline
			%60^{\circ} & $4.57\% $ & $-2.17\% $ & $-36.24\% $ \\
			%80^{\circ} & $3.84\%$ & $-0.20\%$ & $-29.54\%$ \\
			%\end{array}\]
		%\end{table}

\begin{table} [h!]
\caption{\label{area}The relative visual area size shrinking $\delta A$ for the Janis-Newman naked singularity with solution parameter $\gamma = 0.51$ with respect to the corresponding disk region for the Schwarzschild black hole for different image order $k=0,1,2$. We study two inclination angles  $i=60^{\circ}$ and $ i=80^{\circ}$.}
\centering
\small
\begin{tabular}{|l|*{3}{c|}}\hline
\backslashbox{$i$}{$k$}
&\makebox[3em]{0}&\makebox[3em]{1}&\makebox[3em]{2}\\\hline\hline
$60^{\circ}$ & 4.57\% &-2.17\% &-36.24\% \\ \hline
$80^{\circ}$ & 3.84\% &-0.20\% &-29.54\% \\ \hline
\end{tabular}
\vspace{0.5cm}
\end{table}

\section{Radiation from a thin accretion disk around the Janis-Newman naked singularity}

In the previous section we investigated the optical appearance of a thin disk consisting of particles moving on circular equatorial orbits around the Janis-Newman naked singularity. We extend our analysis by considering a physical model of radiation of the disk and studying its image as seen by a distant observer. We adopt the Novikov-Thorne model of a thin accretion disk consisting of anisotropic fluid moving in the equatorial plane \cite{Novikov}, \cite{Page:1974}. The disk hight is negligible compared to its horizontal size, and the disk is stabilized at hydrodynamic equilibrium. The physical properties of the accretion disk are governed by certain structure equations, which follow from the requirement of the conservation of the rest mass, the energy, and the angular momentum of the fluid. Using these conservation laws we can derive the flux of the radiant energy over the disk spanning between the ISCO and a certain radial distance $r$ \cite{Novikov}, \cite{Page:1974}

\begin{equation}\label{F_r}
 F(r)=-\frac{\dot{M}_{0}}{4\pi \sqrt{-g^{(3)}}}\frac{\Omega
_{,r}}{(E-\Omega
L)^{2}}\int_{r_{{\small ISCO}}}^{r}(E-\Omega
L)L_{,r}dr.  \label{F}
\end{equation}

We denote by $g^{(3)}$  the determinant of the induced metric in the equatorial plane, $\dot M_0$ is the mass accretion rate, and $E$, $L$, and $\Omega$ are the energy, the angular momentum, and the angular velocity of the particles moving on a particular circular orbit. For steady-state accretion disk models the accretion rate is independent of time, i.e. it is given by a constant depending on the compact object sourcing the gravitational field. The kinematic quantities depend only on the radius of the orbit, and  they are determined explicitly by the expressions \cite{Harko:2009} for a general static spherically symmetric metric ($\ref{metric_st}$)

\begin{eqnarray}
E&=&-\frac{g_{tt}}{\sqrt{-g_{tt}-g_{\phi\phi}\Omega^2}},    \label{rotE}  \\
L&=&\frac{g_{\phi\phi}\Omega}{\sqrt{-g_{tt}-g_{\phi\phi}\Omega^2}},     \label{rotL}  \\
\Omega&=&\frac{d\phi}{dt}=\sqrt{-\frac{g_{tt,r}}{g_{\phi\phi,r}}}.     \label{rotOmega}
\end{eqnarray}

\begin{figure}[h!]
\centering
    		\includegraphics[width=0.75\textwidth]{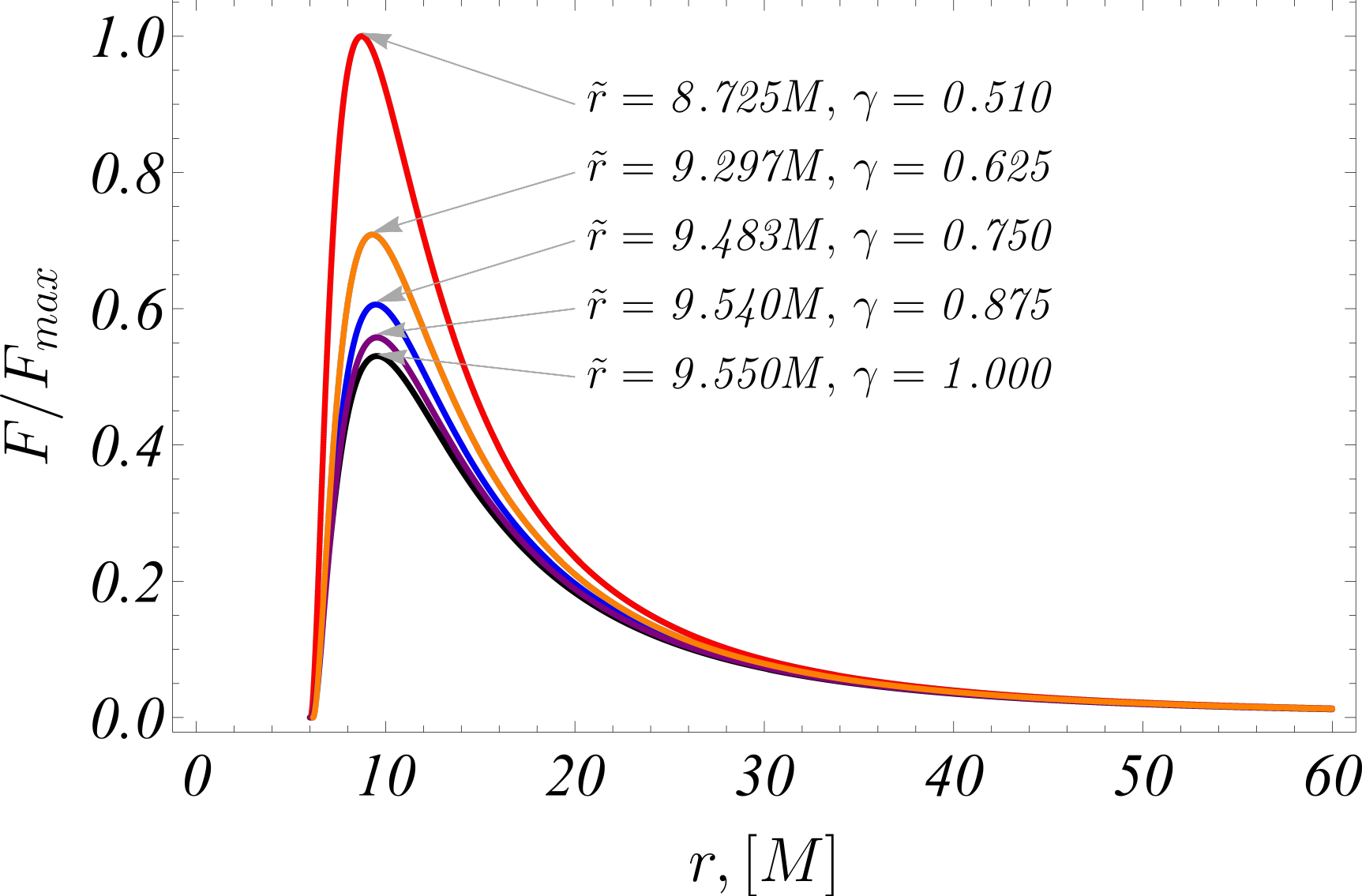}
    \caption{\label{fig:Flux}\small Dependence of the radiation energy flux over the disk on the radial distance for different value of the solution parameter $\gamma$ in the range $\gamma \in (0.5, 1]$.  The position of the flux maximum is denoted by $\tilde r$ and it is specified for each value of $\gamma$ next to the corresponding curve. All the curves are normalized by the maximum of the energy flux for $\gamma=0.51$.}
\end{figure}

We calculate the radiation flux numerically and illustrate its behavior as a function of the radial distance for different values of the solution parameter $\gamma$ (see fig. $\ref{fig:Flux}$). The flux possesses a single maximum,  which grows monotonically when decreasing the value of $\gamma$ in the range $\gamma \in (0.5, 1]$. At the same time its radial position is shifted towards the location of the curvature singularity. All the curves presented in fig. $\ref{fig:Flux}$ are normalized by the maximum of the energy flux for $\gamma = 0.51$, which takes the value $F_{max} = 25.78\times10^{-6}M\dot{M}$, where $M$ is the mass of the compact object. The most substantial part of the radiation is emitted by the disk portion extending up to the radial distance $r=30M$. This gives us a further reason to consider this range as representative for studying the interaction of the accreting matter with the gravitational field of the compact object in the strong field regime.

%accretion rate in F_{max}

The radiation flux will appear deformed to a distant observer, as its apparent intensity $F_{obs}$ in each point of the observer's sky will be modified by a factor depending on the gravitational redshift $z$ \cite{Luminet:1979}

\begin{equation}
F_{obs} = \frac{F}{(1+z)^4}.
\end{equation}

\begin{figure}[h!]
    		\setlength{\tabcolsep}{ 0 pt }{\footnotesize\tt
		\begin{tabular}{ cc}
           \includegraphics[width=0.5\textwidth]{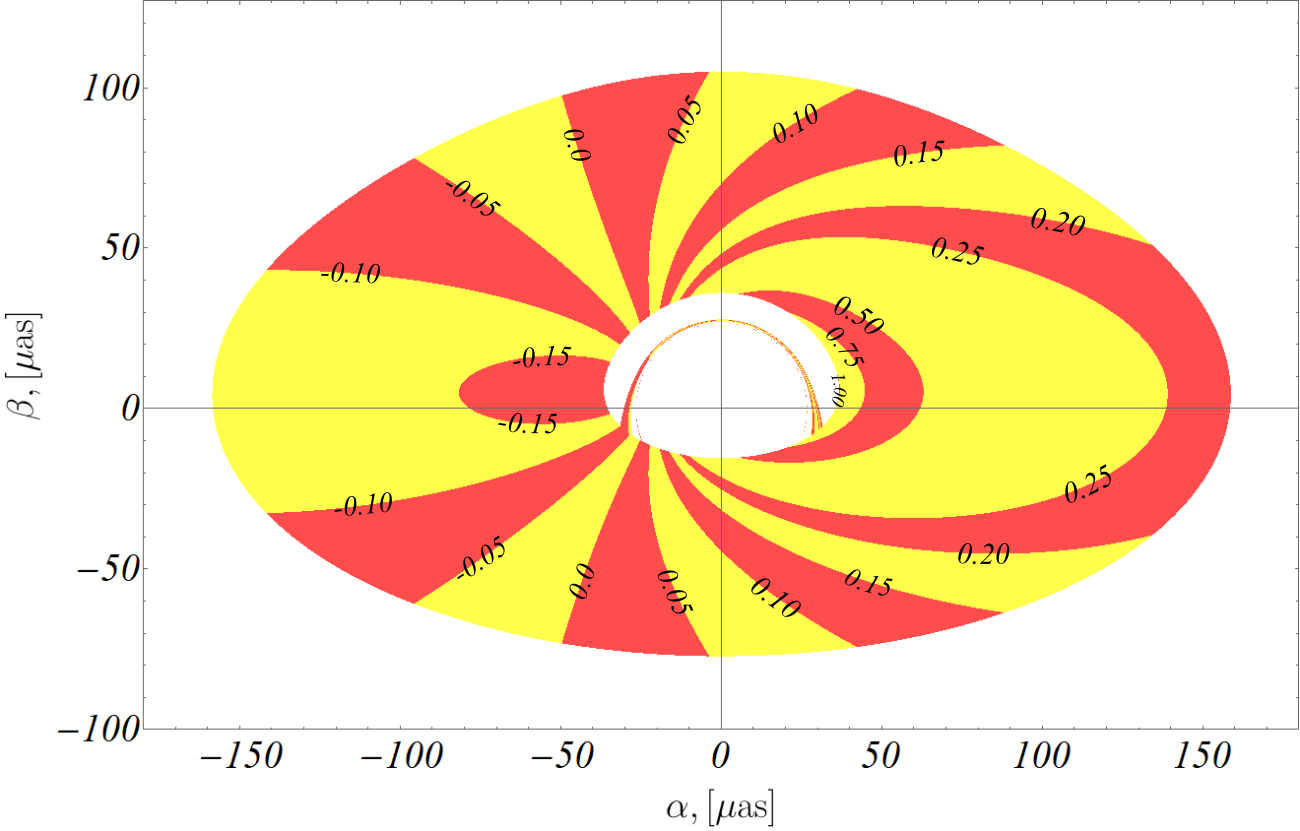}
           \includegraphics[width=0.5\textwidth]{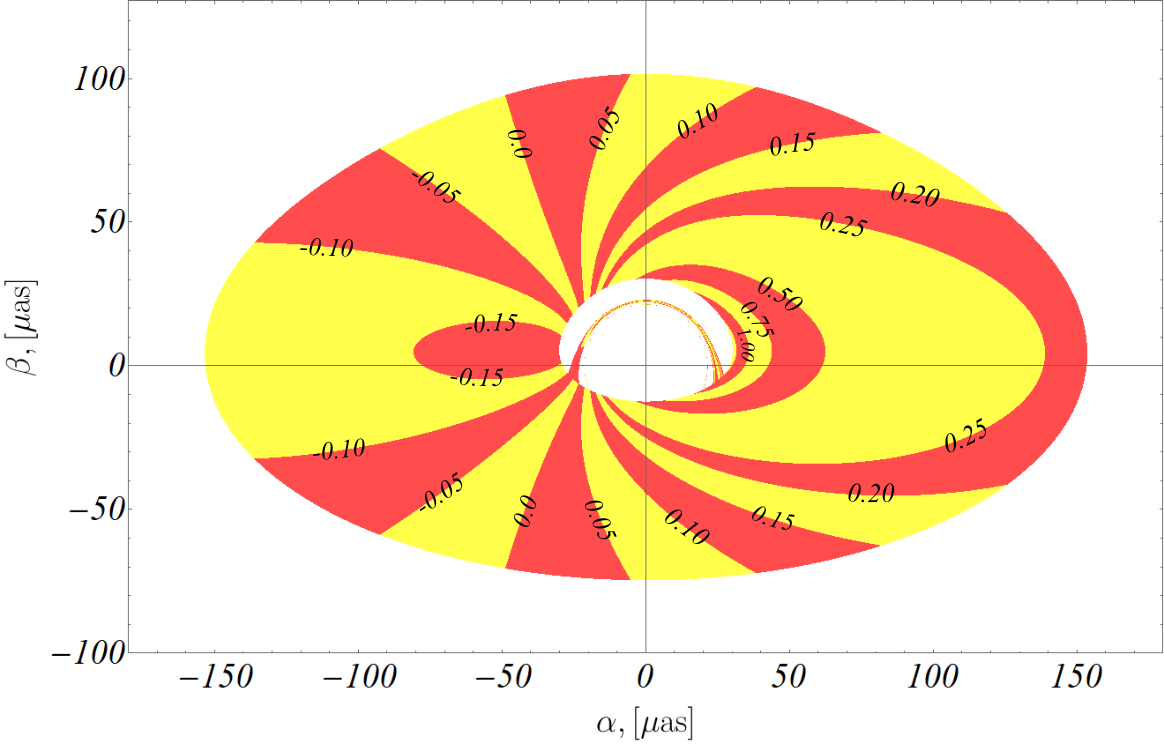}\\[1mm]
			 \hspace{0.4cm} $\gamma=1$, $i=60^\circ$ \hspace{5.0cm}  $\gamma=0.51$, $i=60^\circ$  \\[2mm]
             \includegraphics[width=0.5\textwidth]{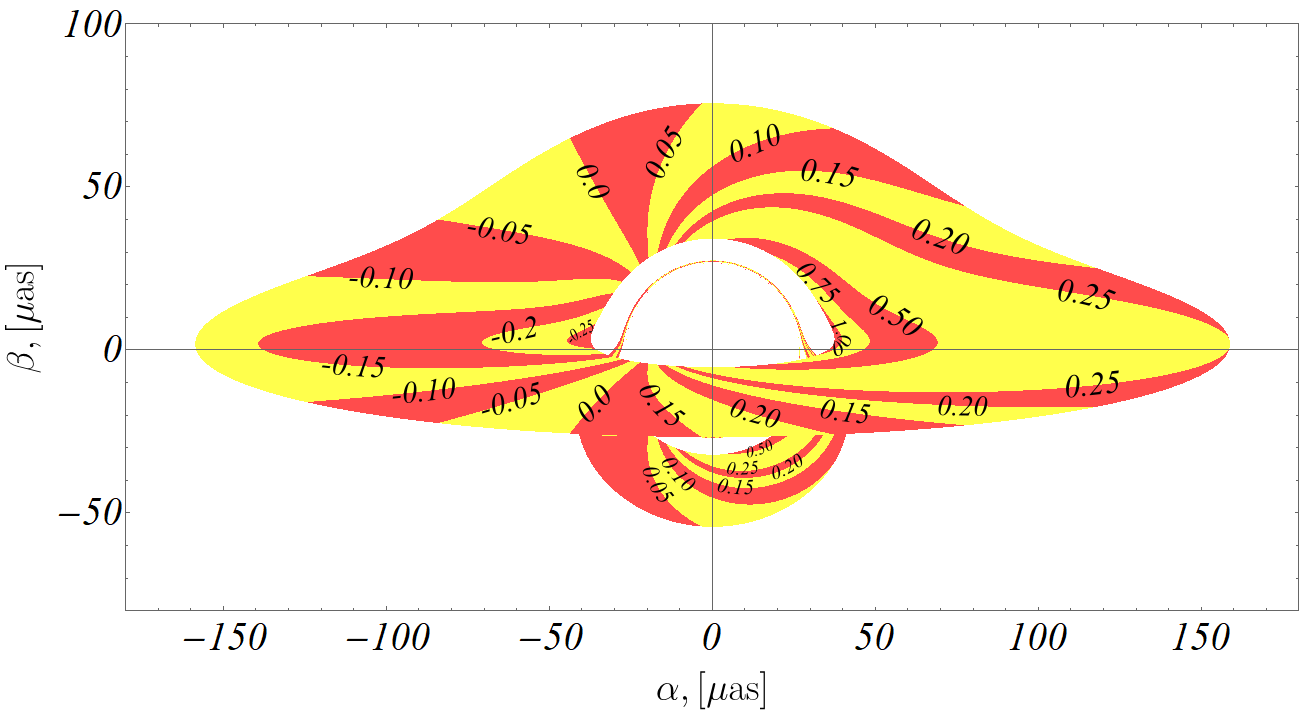}
             \includegraphics[width=0.5\textwidth]{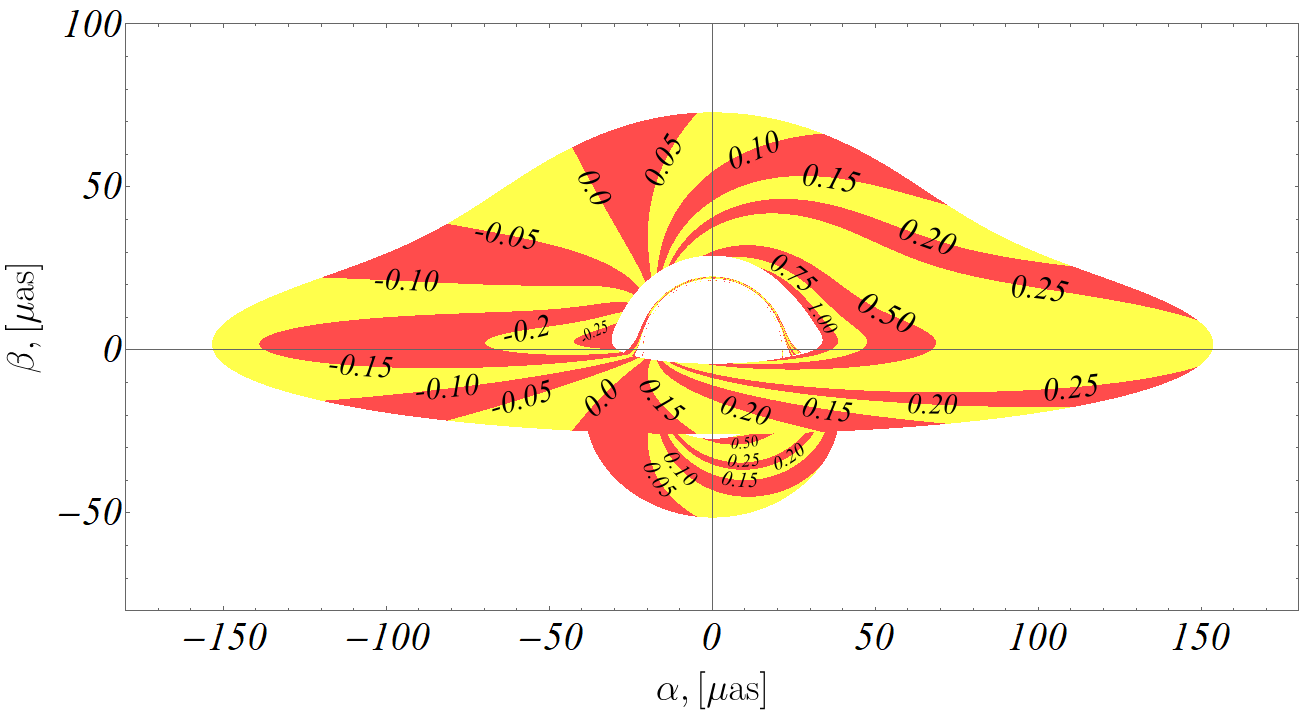} \\[1mm]
             \hspace{0.4cm} $\gamma=1$, $i=80^\circ$ \hspace{5.0cm}  $\gamma=0.51$, $i=80^\circ$
		\end{tabular}}
 \caption{\label{fig:IsoZ}\small Optical appearance of the curves of constant redshift $z$ for the Schwarzschild black hole (left column), and the Janis-Newman naked singularity with solution parameter $\gamma=0.51$ (right column). We present the images for two inclination angles $i=60^\circ$ and $i=80^\circ$. The value of the relevant redshift is specified on each contour.}
\end{figure}

For a general static spherically symmetric metric the redshift factor is given by the expression \cite{Luminet:1979}

\begin{equation}
1+z=\frac{1+\Omega D}{\sqrt{ -g_{tt} - \Omega^2 g_{\phi\phi}}},
\end{equation}
where $D=L/E$ is the impact parameter, which related to the celestial coordinates $\alpha$ and $\beta$ by means of eqs. ($\ref{p_alpha}$). In order to get intuition about the observable flux we investigate its variation as appearing on the observer's sky. In fig. $\ref{fig:IsoZ}$ we present the curves of constant redshift $z$ for the Schwarzschild black hole and the Janis-Newman naked singularity with $\gamma = 0.51$. Both solutions possess very similar observable redshift distributions. A region of negative redshift (blueshift) exists for negative values of the celestial coordinate $\alpha$, which will lead to a region with maximal intensity of the observable radiation flux.

\begin{figure}[h!]
    		\setlength{\tabcolsep}{ 0 pt }{\footnotesize\tt
		\begin{tabular}{ cc}
           \includegraphics[width=0.5\textwidth]{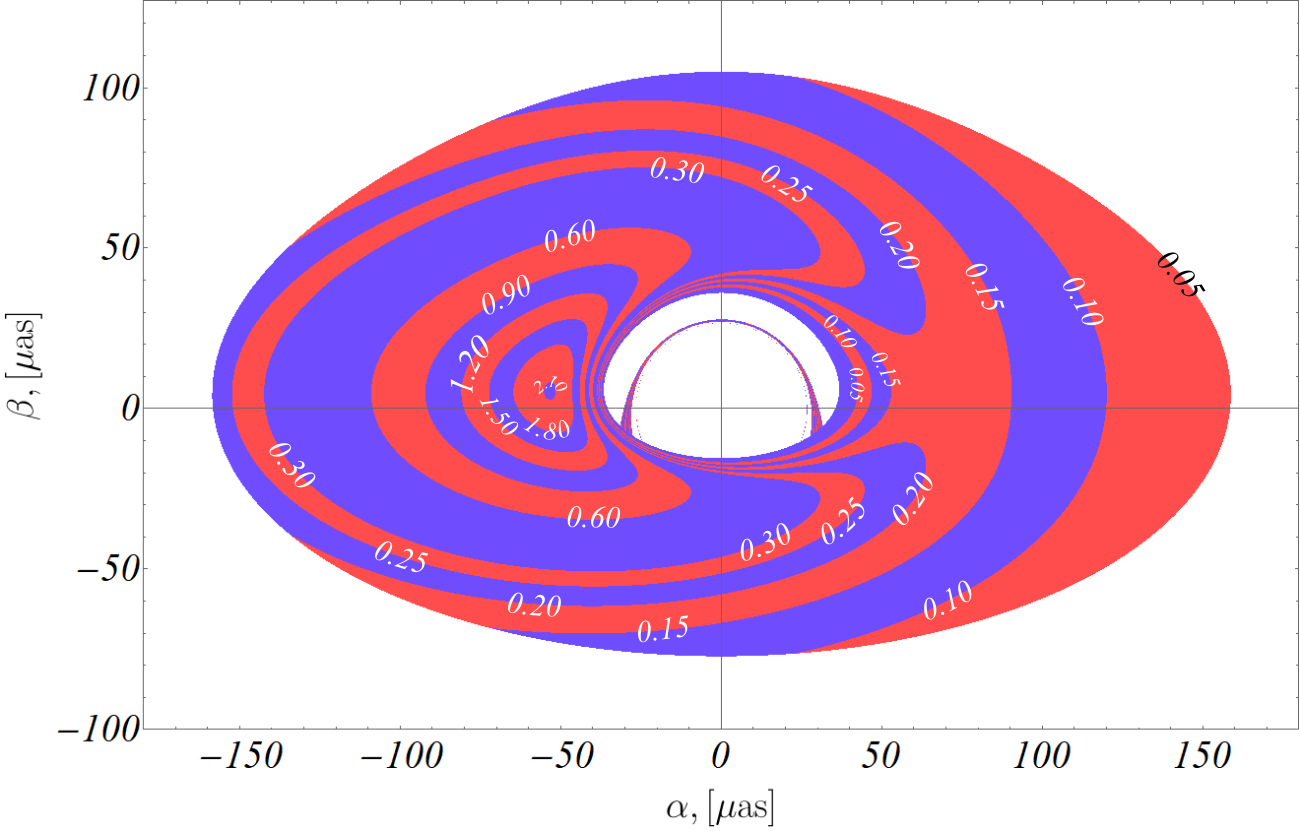}
           \includegraphics[width=0.5\textwidth]{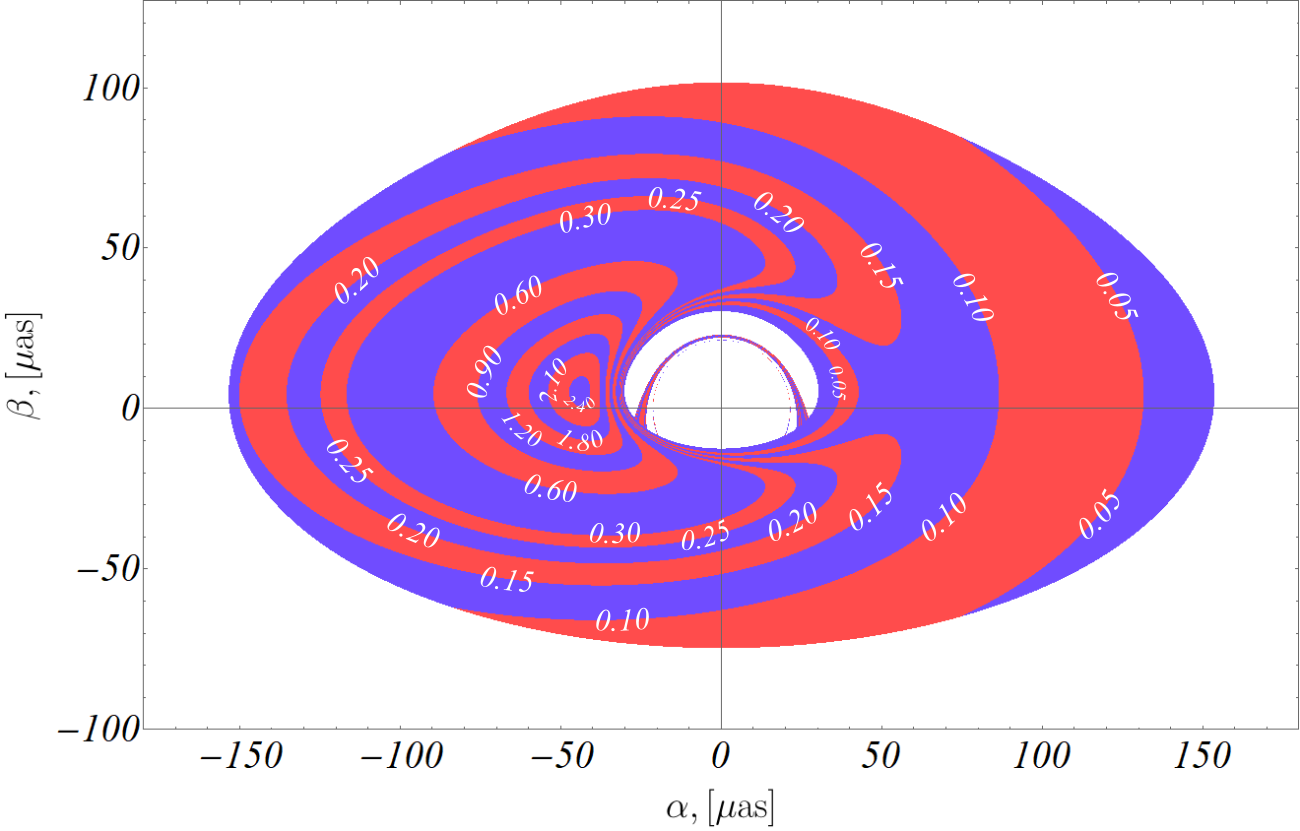}\\[1mm]
			 \hspace{0.4cm} $\gamma=1$, $i=60^\circ$ \hspace{5.0cm}  $\gamma=0.51$, $i=60^\circ$  \\[2mm]
             \includegraphics[width=0.5\textwidth]{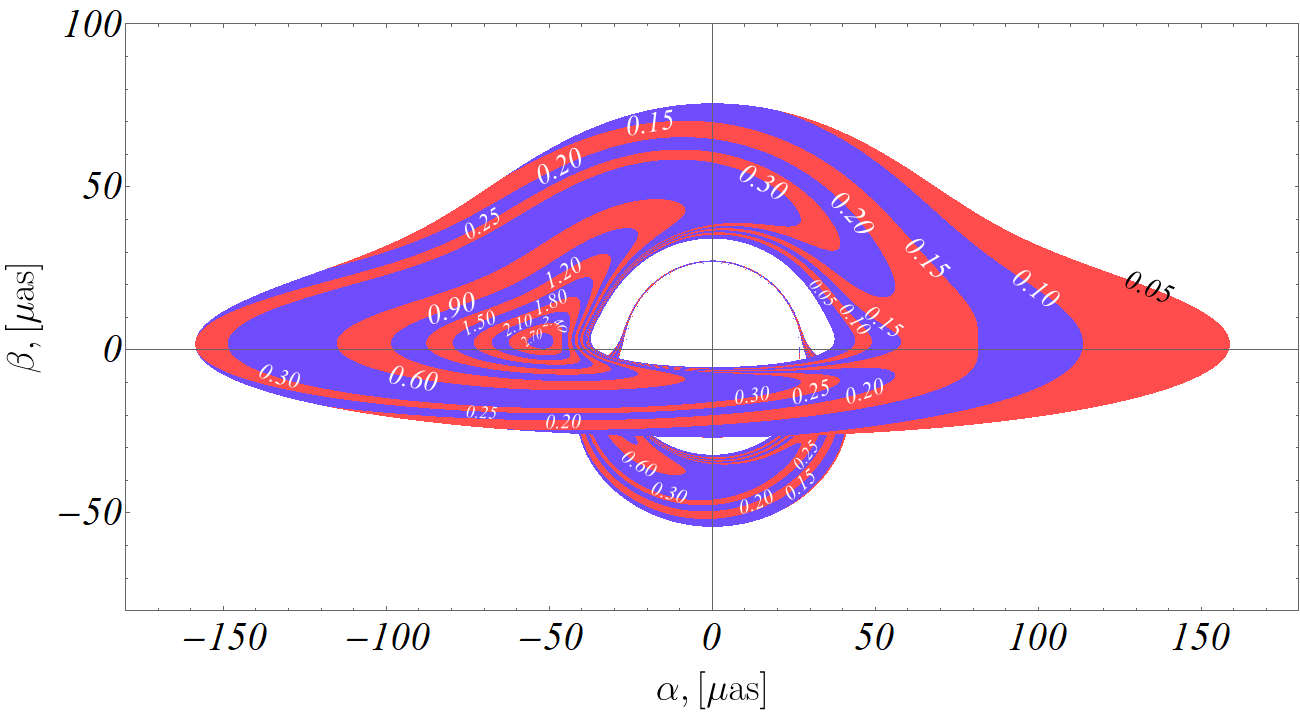}
             \includegraphics[width=0.5\textwidth]{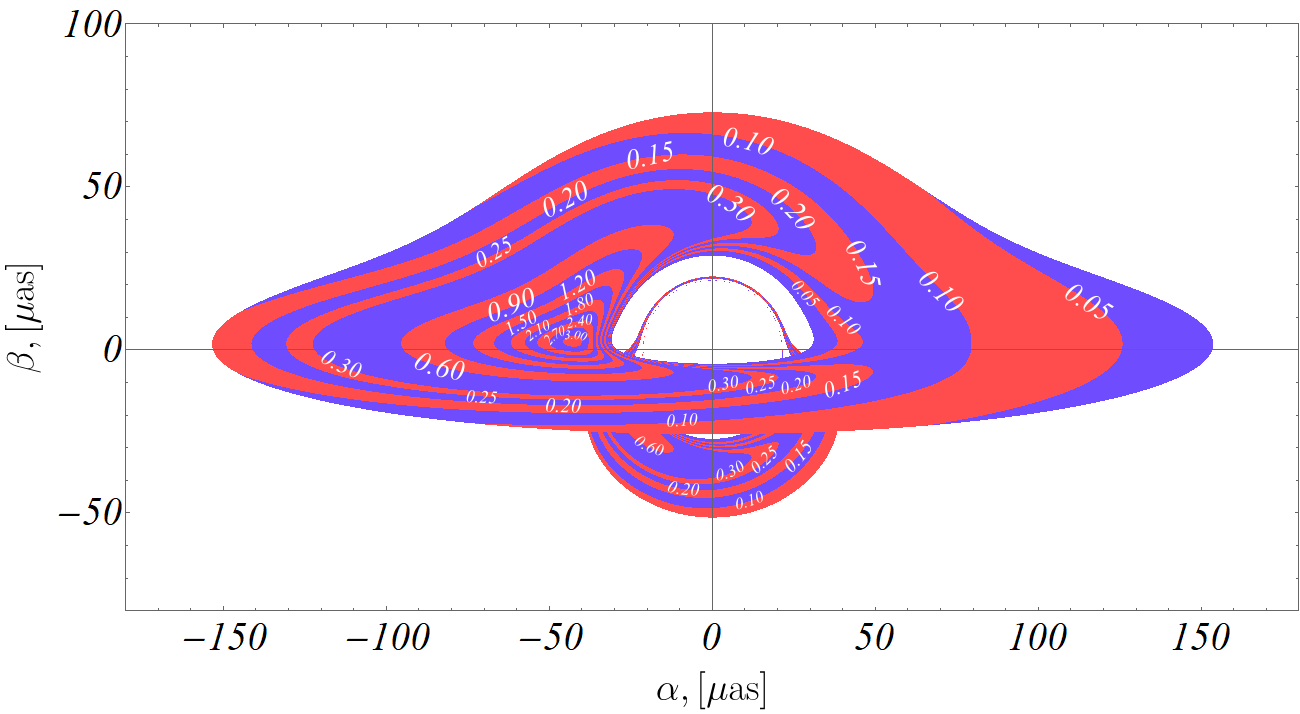} \\[1mm]
             \hspace{0.4cm} $\gamma=1$, $i=80^\circ$ \hspace{5.0cm}  $\gamma=0.51$, $i=80^\circ$
		\end{tabular}}
 \caption{\label{fig:IsoFlux}\small Contour lines of the apparent radiation flux for the Schwarzschild black hole (left column), and the Janis-Newman naked singularity with solution parameter $\gamma=0.51$ (right column). We present the images for two inclination angles $i=60^\circ$ and $i=80^\circ$. The observable flux is normalized by the maximal value of the radial flux distribution for each solution.}
\end{figure}

\begin{figure}[h!]
    		\setlength{\tabcolsep}{ 0 pt }{\footnotesize\tt
		\begin{tabular}{ cc}
           \includegraphics[width=0.5\textwidth]{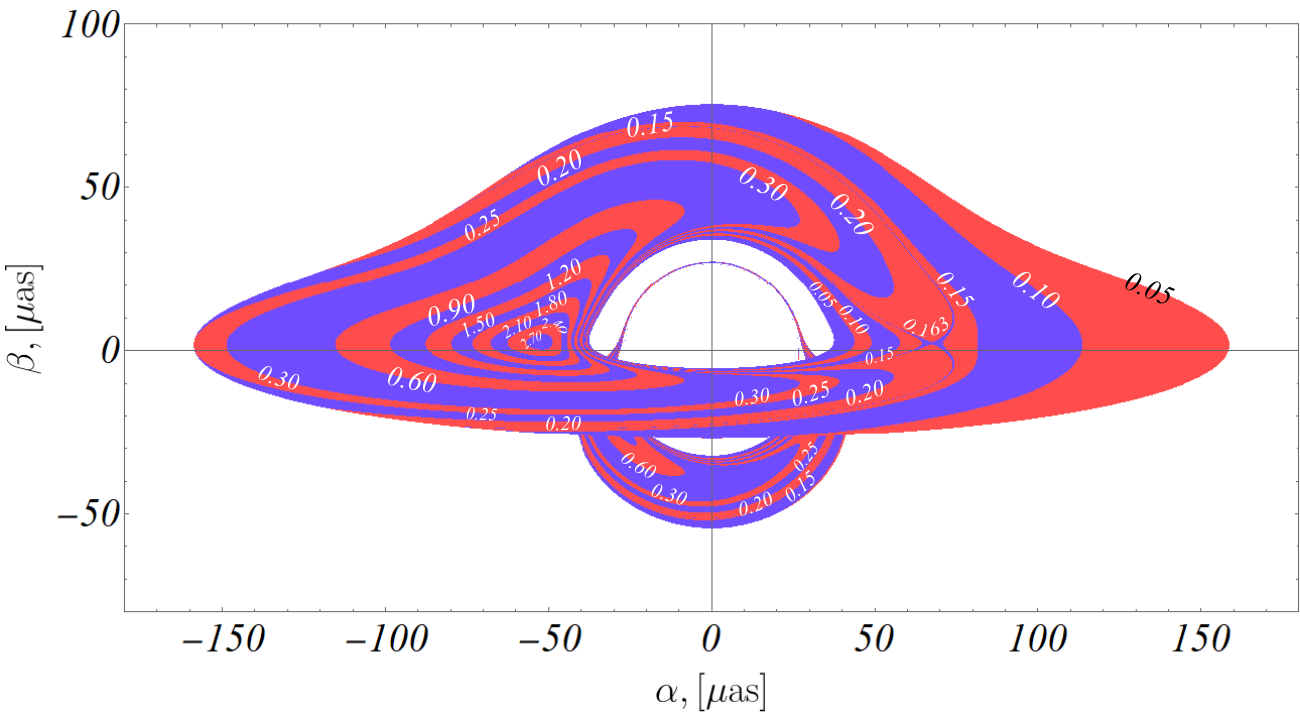}
           \includegraphics[width=0.5\textwidth]{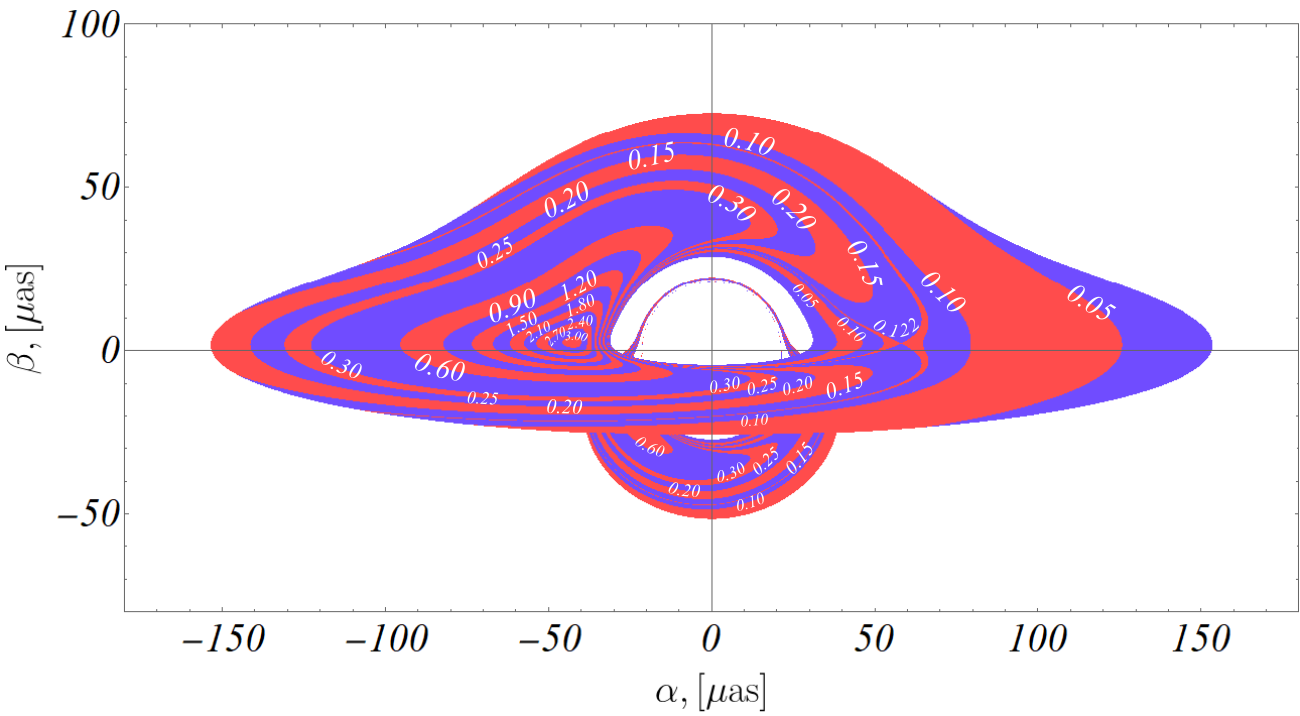}\\[1mm]
			 \end{tabular}}
 \caption{\label{fig:Flux_D}\small Saddle points in the apparent flux distribution for the Schwarzschild black hole (left), and the Janis-Newman naked singularity with solution parameter $\gamma=0.51$ (right). The saddle point for the Janis-Newman solution is located closer to the inner edge of the disk, and corresponds to a  lower value of the relative flux (see main text). The inclination angle is $i=80^\circ$. }
\end{figure}

\begin{figure}[h!]
    		\setlength{\tabcolsep}{ 0 pt }{\footnotesize\tt
		\begin{tabular}{ cc}
           \includegraphics[width=0.5\textwidth]{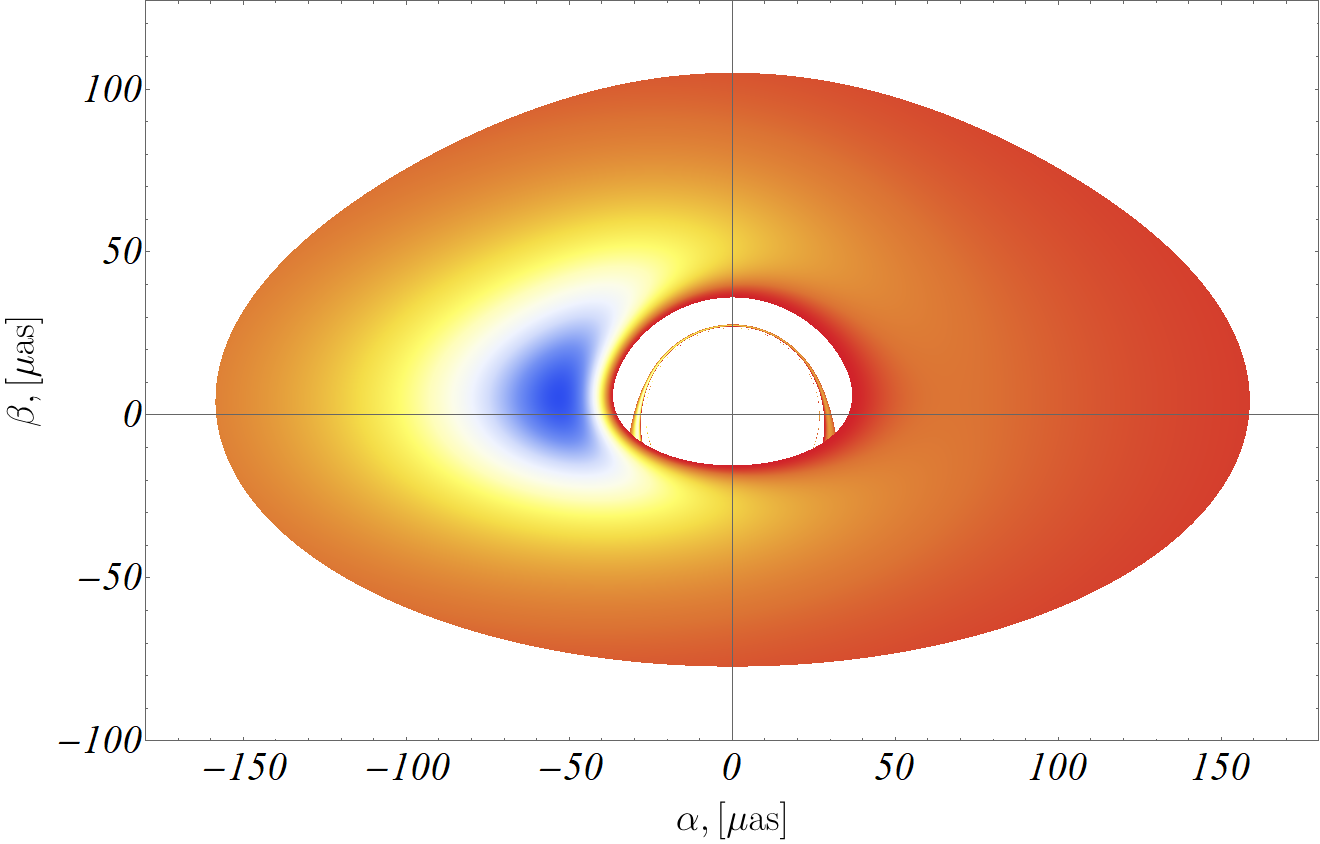}
           \includegraphics[width=0.5\textwidth]{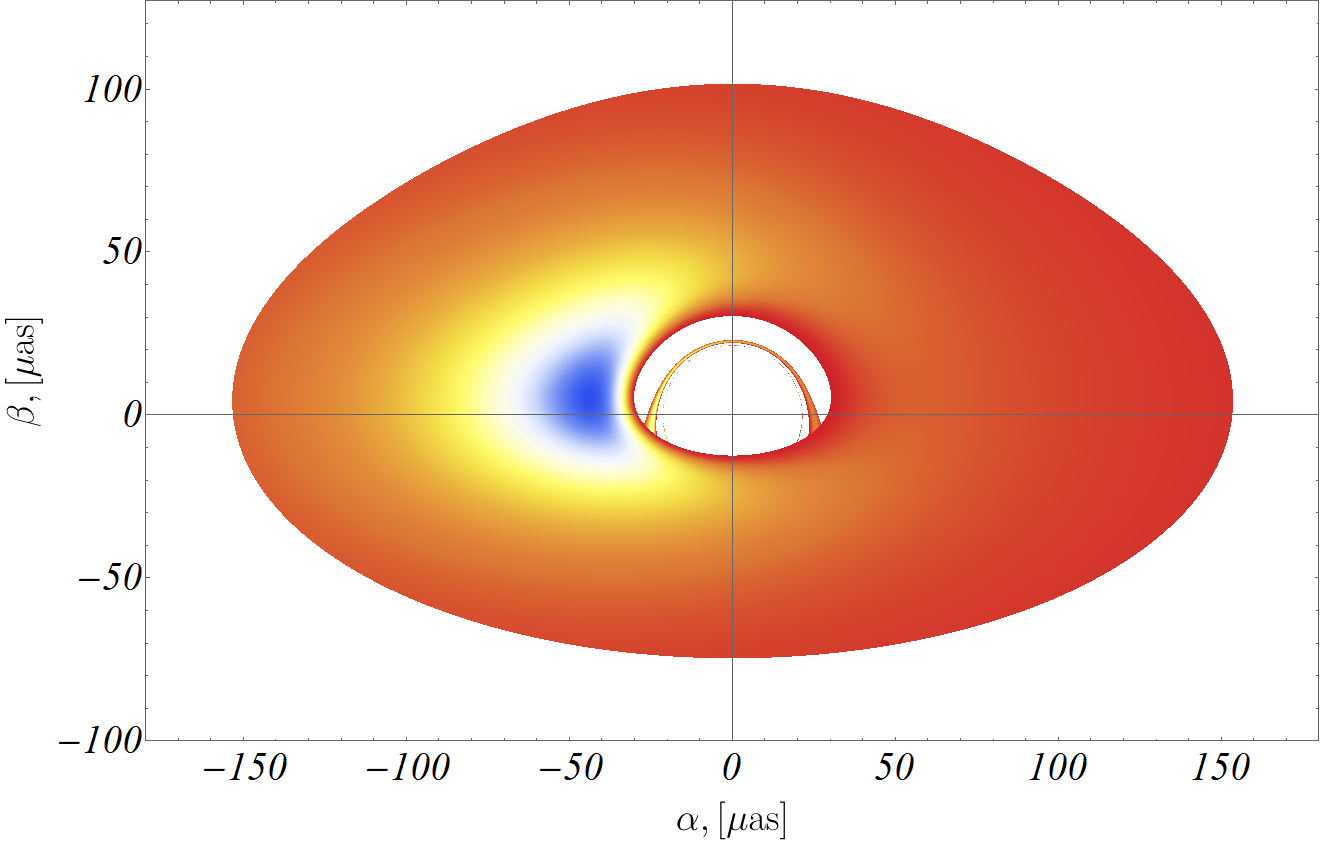}\\[1mm]
			 \hspace{0.4cm} $\gamma=1$, $i=60^\circ$ \hspace{5.0cm}  $\gamma=0.51$, $i=60^\circ$  \\[2mm]
             \includegraphics[width=0.5\textwidth]{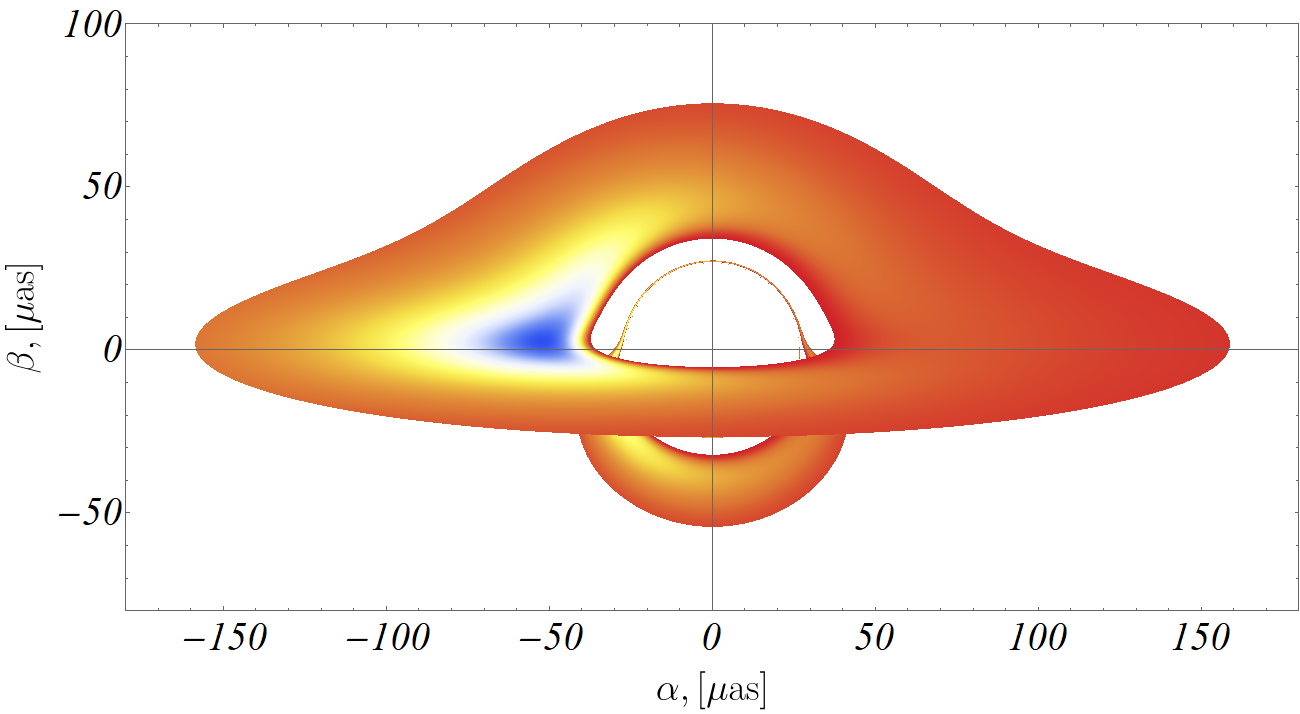}
             \includegraphics[width=0.5\textwidth]{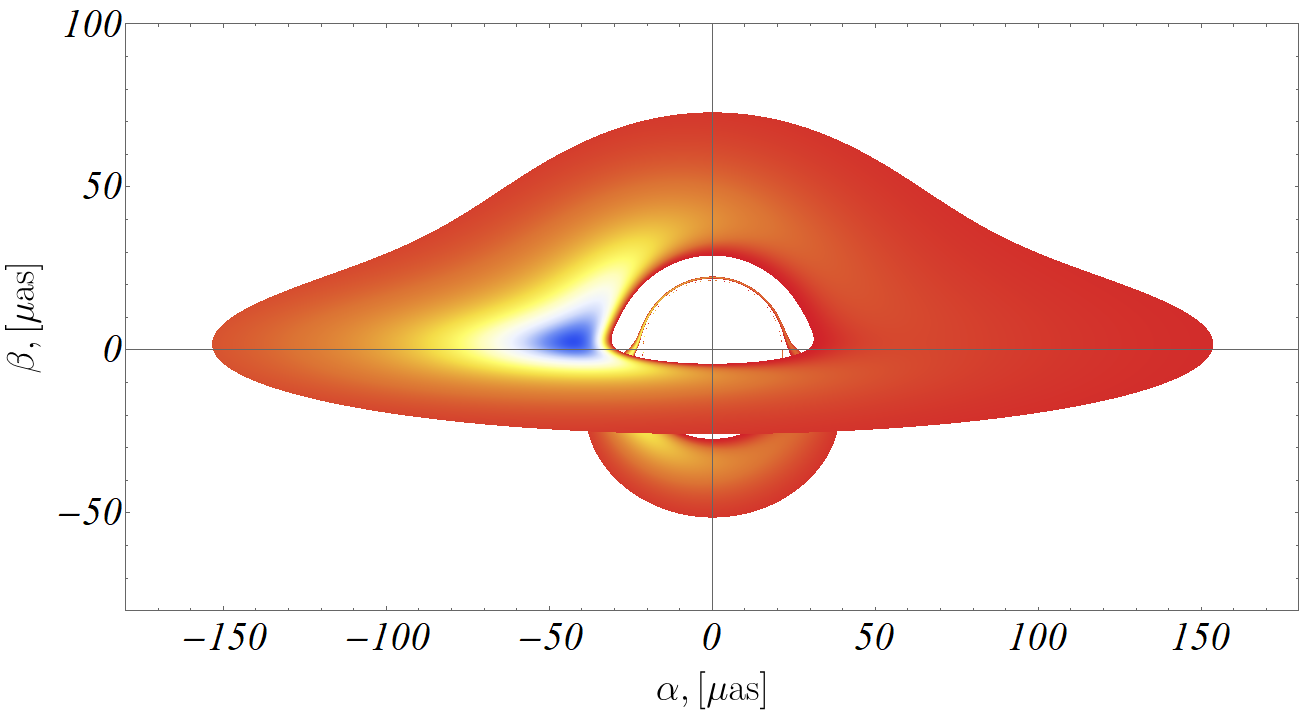} \\[1mm]
             \hspace{0.4cm} $\gamma=1$, $i=80^\circ$ \hspace{5.0cm}  $\gamma=0.51$, $i=80^\circ$
		\end{tabular}}
 \caption{\label{fig:ColorDisk}\small Continuous distribution of the apparent radiation flux for the Schwarzschild black hole (left column), and the Janis-Newman naked singularity with solution parameter $\gamma=0.51$ (right column). We present the images for two inclination angles $i=60^\circ$ and $i=80^\circ$. The flux is normalized by the maximal value of the observable flux for each solution. }
\end{figure}

In figs. $\ref{fig:IsoFlux}$ -$\ref{fig:ColorDisk}$ we study the apparent flux intensity as seen by a distant observer for the Schwarzschild black hole and the Janis-Newman solution with $\gamma = 0.51$. The apparent flux distribution is similar for the two solutions, and only quantitative differences occur. We present two types of images - contour plots, which give the curves of constant flux, and color images, which illustrate the continuous distribution of the apparent radiant energy by associating certain colors to the different intensities. In the contour plots the apparent flux $F_{obs}$ is normalized by the maximum value of the radial flux distribution ($\ref{F_r}$) for the particular solution. As expected the maximum value of the radiation flux appears for negative $\alpha$, where the redshift is negative, and is located near the visual image of the orbit with the appropriate radius, for which the radial distribution of the radiation energy ($\ref{F_r}$) has a maximum. Another characteristic feature of the image is the occurrence of a saddle point in the apparent flux as a function of the celestial coordinates. It is located opposite to the flux maximum with respect to the $\beta$-axis at similar values of the celestial coordinates. In fig. $\ref{fig:Flux_D}$ we illustrate the effect more clearly  by providing a more detailed contour map. For the Schwarzschild solution the saddle point corresponds to the relative flux value $F_{obs}/F_{max} \approx 0.16$, and it is located at the celestial coordinates $(\alpha, \beta) = (67.77, 2.52)$ for the inclination angle $i=80^\circ$. For the Janis-Newman naked singularity the flux value is $F_{obs}/F_{max} \approx 0.12$ at $(\alpha, \beta) = (57.77, 2.82)$ for the same inclination angle. On the other hand the relative flux maximum possesses the value $F_{obs}/F_{max} \approx 2.81$, and location $(\alpha, \beta) = (-52.19, 2.52)$ for the Schwarzschild black hole, and  $F_{obs}/F_{max} \approx 3.11$, $(\alpha, \beta) = (-42.19, 2.52)$ for the Janis-Newman solution. We see that for the Janis-Newman solution the maximal value of the observable flux is higher than for the Schwarzschild black hole, and both extrema are shifted more closely to the inner edge of the disk.

We illustrate the continuous distribution of the apparent radiation flux in fig. $\ref{fig:ColorDisk}$. Contrary to the contour plots, in this case the flux is normalized by the maximal value of the observable flux for each solution, so that its values belong to the interval $[0,1]$. We define a color map associating continuously a certain color of the spectrum from red to blue to each value in the interval $[0,1]$. Red corresponds to the lowest flux, while blue denotes the maximum flux value. From the contour plots we see that for the Schwarzschild black hole the flux is more evenly distributed, and grows more gradually toward its peak than for the Janis-Newman solution. For the Janis-Newman naked singularity the flux possesses lower values for positive $\alpha$ and in this region grows more slowly than for the Schwarzschild black hole. Yet, in the vicinity of the flux maximum, it increases more steeply, and reaches a higher value at the peak. Therefore, for this solution the apparent emitted radiation is more concentrated around the observable location of its maximum. Since all the visualized flux values are normalized by the maximum value of the observable flux, the disk for the Janis-Newman solution appears redder than that of for the Schwarzschild black hole. In this case the region for positive $\alpha$ is characterized by lower apparent flux values, while the maximal value of the apparent flux is higher. This lead to further decreasing of the values of the normalized flux.

\section{Conclusion}

In this paper we study the optical appearance of a thin accretion disk surrounding the Janis-Newman-Winicour naked singularity and its observable radiation in the framework of the Novikov-Thorne model. We concentrate on the case when the Janis-Newman solution possesses a photon sphere, which ensures that its lensing properties are similar to those of the Schwarzschild black hole. In particular, it gives rise to a shadow. We aim at comparing the images produced by the accretion disk around the Janis-Newman naked singularity and the Schwarzschild black hole, and deducing observable effects, which can differentiate between the two spacetimes, and can be useful in the interpretation of future experimental data. We observe qualitatively very similar features in both cases, so the two solutions can be distinguished only by quantitative measurements. The Janis-Newman naked singularity leads to a smaller direct image of the accretion disk, with a higher maximum of the radiation flux. Its observable radiation is more unevenly distributed across the disk, as it exhibits lower values in its periphery, and concentrates in its largest portion in a smaller neighbourhood of the emission peak. In a future work we will extend our analysis to the case of the strongly naked Janis-Newman singularity when no photon sphere is present, and the spacetime can be qualitatively distinguished by the Schwarzschild black hole by accretion disk observations.

\section*{Acknowledgments}
The authors gratefully acknowledge support by the Sofia University Research Fund under Grant  No. 80-10-129/15.04.2019.
P.N. is supported by the Bulgarian NSF Grant DM 18/3 and the DFG Research Training Group 1620 “Models of Gravity”. S. Y. acknowledges financial support by the Bulgarian NSF Grant No. KP-06-H28/7. Networking support by the COST Action CA16104 is also gratefully acknowledged.

\end{document}